\documentclass{elsart}

\usepackage{graphicx}
\usepackage{color}
\usepackage{multicol}
\usepackage{multirow}
\usepackage{hyperref}

\usepackage{syntax}
\shortverb{\¦} 

\newcommand{\ROOT}{\href{http://root.cern.ch}{
ROOT
}}
\newcommand{\Cls}[1]{
\href{http://root.cern.ch/root/htmldoc/#1.html}{
\textsf{\textsl{\color[rgb]{0,0,0.8}#1}}}}               
\newcommand{\Cmp}[1]{\textsf{\color[rgb]{0.6,0,0}#1}}    
\newcommand{\NmSp}[1]{\texttt{\color[rgb]{0,0.5,0}#1}}   
\newcommand{\Mtd}[1]{\textsl{\color[rgb]{0,0,0.8}#1}}    
\newcommand{\Alg}[1]{\textrm{\color{blue}#1}}   
\newcommand{\Pkg}[1]{\textsf{#1}}               
\newcommand{\Lib}[1]{\texttt{\color[rgb]{0,0.5,0.5}#1}}  

\newcommand{\ACLiC}{\Cmp{ACLiC}}

\newcounter{bla}
\newenvironment{refnummer}{%
\list{[\arabic{bla}]}%
{\usecounter{bla}%
 \setlength{\itemindent}{0pt}%
 \setlength{\topsep}{0pt}%
 \setlength{\itemsep}{0pt}%
 \setlength{\labelsep}{2pt}%
 \setlength{\listparindent}{0pt}%
 \settowidth{\labelwidth}{[9]}%
 \setlength{\leftmargin}{\labelwidth}%
 \addtolength{\leftmargin}{\labelsep}%
 \setlength{\rightmargin}{0pt}}}
 {\endlist}

\begin{document}
\begin{frontmatter}

\title{ROOT --- A C++ Framework for Petabyte Data Storage,
    Statistical Analysis and Visualization}

\author{I.~Antcheva},
\author{M.~Ballintijn},
\author[a]{B.~Bellenot},
\author{M.~Biskup},
\author[a]{R.~Brun\thanksref{author-rb}},
\author{N.~Buncic},
\author[b]{Ph.~Canal},
\author[c]{D.~Casadei}, 
\author[a]{O.~Couet},
\author[d]{V.~Fine},
\author{L.~Franco},
\author[a]{G.~Ganis},
\author[a]{A.~Gheata},
\author[a]{D.~Gonzalez~Maline},
\author{M.~Goto},
\author[a]{J.~Iwaszkiewicz},
\author{A.~Kreshuk},
\author{D.~Marcos~Segura},
\author{R.~Maunder},
\author[a]{L.~Moneta},
\author[a]{A.~Naumann}, 
\author{E.~Offermann},
\author{V.~Onuchin},
\author[b]{S.~Panacek},
\author[a]{F.~Rademakers}, 
\author[b]{P.~Russo},
\author[a]{M.~Tadel}

\thanks[author-rb]{Ren\'e Brun, \url{rene.brun@cern.ch}}

\address[a]{CERN, Geneva, Switzerland}
\address[b]{Fermilab, Batavia, IL, USA}
\address[c]{New York University, NY, USA}
\address[d]{Brookhaven National Lab, Upton, NY, USA}

\begin{abstract}  
  ROOT is an object-oriented C++ framework conceived in the
  high-energy physics (HEP) community, designed for storing and
  analyzing petabytes of data in an efficient way.  Any instance of a
  C++ class can be stored into a ROOT file in a machine-independent
  compressed binary format.  In ROOT the TTree object container is
  optimized for statistical data analysis over very large data sets by
  using vertical data storage techniques. These containers can span a
  large number of files on local disks, the web, or a number of
  different shared file systems. In order to analyze this data, the
  user can chose out of a wide set of mathematical and statistical
  functions, including linear algebra classes, numerical algorithms
  such as integration and minimization, and various methods for
  performing regression analysis (fitting). In particular, the RooFit
  package allows the user to perform complex data modeling and fitting
  while the RooStats library provides abstractions and implementations
  for advanced statistical tools. Multivariate classification methods
  based on machine learning techniques are available via the TMVA
  package. A central piece in these analysis tools are the histogram
  classes which provide binning of one- and multi-dimensional
  data. Results can be saved in high-quality graphical formats like
  Postscript and PDF or in bitmap formats like JPG or GIF. The result
  can also be stored into ROOT macros that allow a full recreation and
  rework of the graphics. Users typically create their analysis macros
  step by step, making use of the interactive C++ interpreter CINT,
  while running over small data samples.  Once the development is
  finished, they can run these macros at full compiled speed over
  large data sets, using on-the-fly compilation, or by creating a
  stand-alone batch program.  Finally, if processing farms are
  available, the user can reduce the execution time of intrinsically
  parallel tasks -- e.g.\ data mining in HEP -- by using PROOF, which
  will take care of optimally distributing the work over the available
  resources in a transparent way.
  
\begin{flushleft}
PACS: 00;  07; 05
\end{flushleft}

\begin{keyword}
C++; object-oriented; framework; interpreter; data storage; data analysis; visualization
\end{keyword}

\end{abstract}

\end{frontmatter}

{\bf PROGRAM SUMMARY}

\begin{small}
\noindent
{\em Manuscript Title:} ROOT --- A C++ Framework for Petabyte Data Storage,
    Statistical Analysis and Visualization   \\
{\em Authors:} Ilka Antcheva, Maarten Ballintijn, Bertrand Bellenot, Marek Biskup, Ren\'e Brun, Nenad Buncic, Philippe Canal, Olivier Couet, Valeri Fine, Leandro Franco, Gerardo Ganis, Andrei Gheata, David Gonzalez Maline, Masaharu Goto, Jan Iwaszkiewicz, Anna Kreshuk, Diego Marcos Segura, Richard Maunder, Lorenzo Moneta, Axel Naumann, Edmond Offermann, Valeri Onuchin, Susan Panacek, Fons Rademakers, Paul Russo, Matevz Tadel\footnote{For an up-to-date
author list see: \url{http://root.cern.ch/drupal/content/root-development-team} and
\url{http://root.cern.ch/drupal/content/former-root-developers}}                 \\
{\em Program Title:} ROOT                    \\
{\em Journal Reference:}                     \\
{\em Catalogue identifier:}                  \\
{\em Licensing provisions:} LGPL             \\
{\em Programming language:} C++              \\
{\em Computer:} Intel i386, Intel x86-64, Motorola PPC, Sun Sparc, HP PA-RISC  \\
{\em Operating system:} GNU/Linux, Windows XP/Vista, Mac OS X, FreeBSD,
    OpenBSD, Solaris, HP-UX, AIX             \\
{\em RAM:} $>$55 Mbytes                      \\
{\em Number of processors used:} $>=$ 1      \\
{\em Keywords:} C++, object-oriented, framework, interpreter, data
    storage, data analysis, visualization    \\ 
{\em PACS:} 00,  07, 05                      \\
{\em Classification:}  4, 9, 11.9, 14        \\

{\em Nature of problem:} Storage, analysis and visualization of scientific data\\
   \\
{\em Solution method:} Object store, wide range of analysis algorithms
   and visualization methods \\ 
{\em Running time:} depending on the data size and complexity of
   analysis algorithms\\ 
   \\
{\em References:}
\begin{refnummer}
\item \url{http://root.cern.ch}
\end{refnummer}
\end{small}

\newpage

\hspace{1pc}
{\bf LONG WRITE-UP}


\section{Introduction}

 {\ROOT} is a cross-platform C++ framework for data processing, created at CERN\footnote{%
 European Organization for Nuclear Research, Geneva, Switzerland.}.  Every day,
 thousands of physicists use {\ROOT} based applications to analyze and
 visualize their data.

 The {\ROOT} project was started in 1995 by Ren\'e Brun and Fons Rademakers \cite{root96}.
 It started as a private project and grew to be the officially supported LHC analysis toolkit.
 It is currently developed by a small team with members from several
 laboratories. {\ROOT} benefits from a considerable amount of user
 contributions, both from inside and outside science. This write-up
 focuses on the current status of {\ROOT}, as of version 5.24.00.

 A typical application developed for HEP research (more details in section
 \S\ref{sec-uses} and figure~\ref{fig-typical-use}) is used
 to process both real and simulated data, consisting of many
 \emph{events} having the same data structure and assumed to be
 statistically independent\footnote{Such independence is very
   important from the computing point of view, because it allows to gain
   the maximum speed-up by distributing subsets of the data to parallel
   analysis nodes.}.  In addition, complementary information is also
 needed to analyze the data, for example detector parameters (geometry,
 read-out powering and configuration, magnetic field maps, etc.) or
 input settings of the simulation engines.  Such values do not
 change at the event scale.  Rather, they have a slower evolution
 that defines a much coarser granularity: a \emph{run} is defined by a
 set of events with constant settings\footnote{In real life, few of
   the auxiliary parameters may be allowed to vary inside a run.  Hence,
   they define smaller blocks that are intermediate between the event
   scale and the run granularity.}.


\subsection{Discovering ROOT}\label{sec-discover}

\begin{figure}[t]
  \centering
  \includegraphics[scale=0.9]{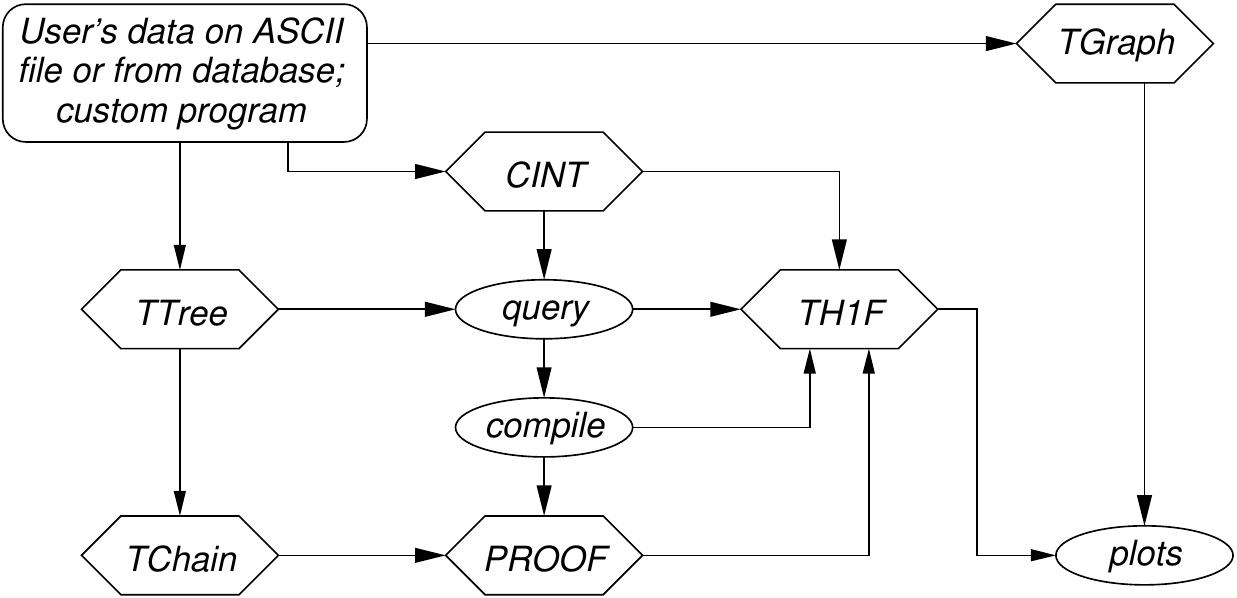}
  \caption{Most frequent approach to start using {\ROOT}.}\label{fig-discovering}
\end{figure}

 To introduce the {\ROOT} framework, one may follow the typical approach
 of new users and its large collection of libraries and tools, with the help of the
 sketch in figure~\ref{fig-discovering}. For a comprehensive description
 of ROOT's features see the User's Guide\cite{usersguide}.

 Newcomers often start from their own analysis program, running over
 their data (usually stored in ASCII format or accessed through a
 relational database engine). They simply look for a library to
 produce graphs to visualize their histograms.
 They start by playing with the {\ROOT} class \Cls{TGraph}, which can be
 used to display a set of $(x,y)$ points including errors.

 The next step is to use {\ROOT} histograms (whose base class is
 \Cls{TH1}; see \S\ref{sec-histo} for more details) instead, and let
 the \Cls{TH1::}\Mtd{Draw()} method produce the plots. {\ROOT}
 histograms can be used for binning a data set and to estimate its
 density.  They have a number of useful properties, allowing the user
 to manipulate them, to obtain statistical information about the
 underlying data, and to perform fits without caring about the plots
 --- they will redraw themselves whenever changes are applied.

 Especially during the early phases, when the data analysis program
 changes quite often, the users may find the interactive C++
 interpreter \Cmp{CINT} embedded in {\ROOT} very useful.  Developing
 programs with the help of an interpreter speeds up the typical
 iterative approach to data analysis by removing the additional
 compile and link steps.  Of course, if the logic of the application
 is already well known, one may prefer to develop the program in a
 more structured way, relying on the compiler in the usual way.

 The most common task for data access in HEP is the
 selective, sparse scanning of data. Traditional RDBMS-like horizontal data partitioning
 does not allow for efficient sparse reading, with the exception of indices. Instead, ROOT uses vertical
 data partitioning of arbitrary user-defined objects, implemented in its \Cls{TTree} class.

 \Cls{TTree}s are partitioned into branches. During reading each branch can be accessed
 independently.  A \Cls{TBranch} stores consecutive objects or data members of
 a class or other \Cls{TBranch}es.  By
 default, all branches stored in a \Cls{TTree} are written into
 separate buffers in a file, so that iterating over the data
 stored in a branch requires only the reading of these associated buffers.
  \Cls{TTree}s can span multiple {\ROOT} files. A {\ROOT} file is very
 similar to a file system, allowing for further internal organization
 using directories.  For example, the main data set could be stored
 into a single \Cls{TTree}, whereas summary information (in the form
 of histograms) resides in separate directories in the same
 \Cls{TFile}.

 If the data volume grows, the user can choose to split the
 \Cls{TTree} instance among several \Cls{TFile} instances.  Later,
 when accessing data, they can all be chained into a single logical
 entity, a \Cls{TChain}, making accessing several files almost
 transparent.  Because a \Cls{TChain} inherits from a \Cls{TTree}, it
 provides the same benefits in terms of optimized data access, even
 though the data are distributed among different files.

 The quickest way to develop the user's analysis program is
 creating {\ROOT} macros step by step using \Cmp{CINT}. Once the
 development phase has ended, performance becomes paramount.
 The first obvious optimization step is to convert the application
 into a compiled program.  Still, one does not need to abandon the use
 of the interpreter: the most efficient way to work with {\ROOT} is to
 consider the interpreter as the ``glue'' which binds together the
 compiled pieces of code that perform most of the intensive
 computation.  Actually, this is less difficult than it appears:
 \Cmp{CINT} macros can be compiled during the interactive session by
 {\ACLiC} (\S\ref{sec-aclic}), to gain the full speed of compiled code
 and the reliability of the full C++ compiler (\Cmp{CINT} has 
 e.g.~limited support of C++ templates).  In general, interpreted code may call
 compiled code and vice versa (more details on \S\ref{sec-bindings}).
 Finally, if a multi-core machine or a computing farm is
 available, \Cmp{PROOF} (\S\ref{sec-parallel}) provides a way to make
 full use of the inherent event parallelism of independent HEP events
 by taking care of distributing the analysis over all available CPU's
 and disks in a transparent way.


\subsection{Typical Uses of ROOT}\label{sec-uses}

\begin{figure}[t!]
  \centering
  \includegraphics[scale=1.0]{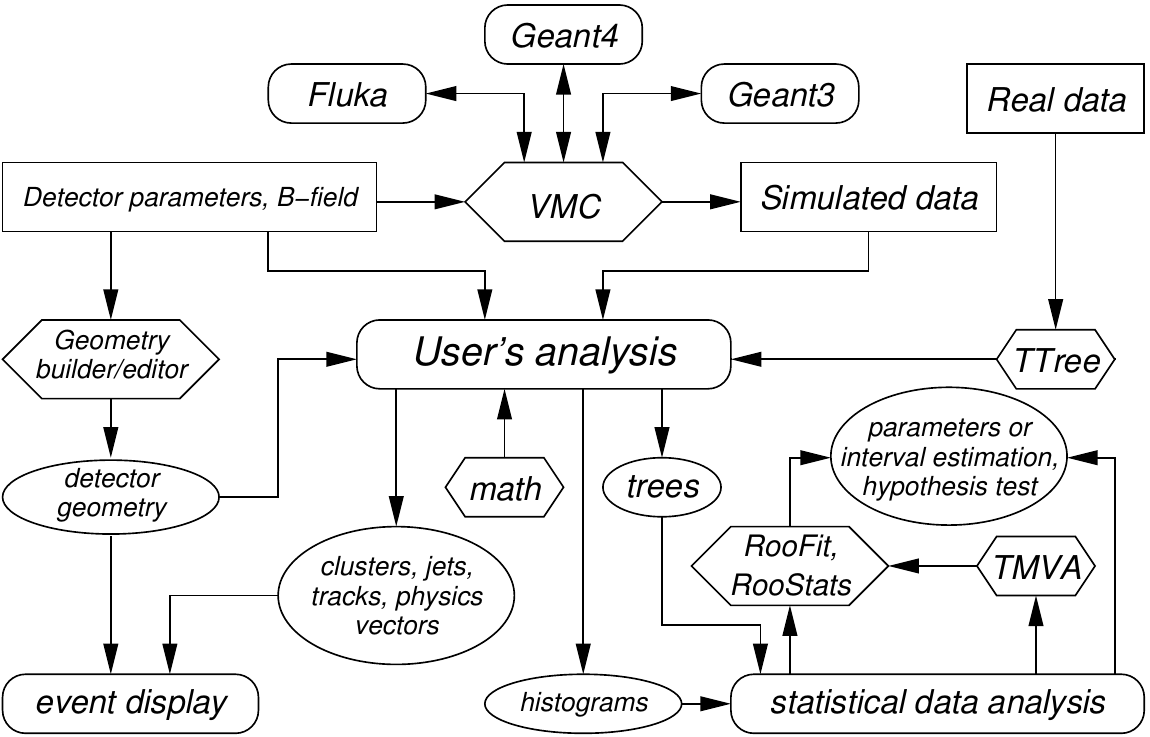}
  \caption{Example of typical usage of {\ROOT}.}\label{fig-typical-use}
\end{figure}

 Figure \ref{fig-typical-use} shows most of the features that a {\ROOT}
 application can have.  Of course, a single application rarely has all
 of them: for example, its focus could be on the detector simulation or on
 the data analysis, but not both.

 {\ROOT} provides the Virtual Monte-Carlo (\Cmp{VMC}) interface
 (\S\ref{sec-simulation}) to the most important HEP simulation engines, like
 \Pkg{Geant4} \cite{geant4} (C++), \Pkg{Geant3} \cite{geant3},
 \Pkg{Fluka} \cite{fluka} (FORTRAN), to simulate
 the passage of particles through matter and their propagation in a
 magnetic field.  The \Cmp{VMC} interface allows the user to build an
 application that simulates the behavior of a particle detector,
 with the freedom to switch between different simulation engines.  
 Comparing the results of different simulation engines allows to 
 estimate systematic simulation uncertainties.

 Usually, most {\ROOT} users develop programs to perform statistical
 data analysis (see also \S\ref{sec-math-stat}) of binned (histograms) or un-binned
 data (\Cls{TTree} variables).  The
 \Pkg{TMVA} package\footnote{\url{http://tmva.sourceforge.net/}}
 (\S\ref{sec-math-stat}) can be used for event classification to
 discriminate between signal and background.  Various methods exist
 for performing the best fits of the selected data to theoretical
 models.

 {\ROOT} can also be used to develop an event display \S\ref{sec-graphics}.
 An event display is an application that provides detector geometry
 visualization, views of hits\footnote{In the HEP jargon, a ``hit'' is
   a localized energy deposition that is detected by the read-out
   electronics.} and clusters of hits used to build calorimeter
 jets\footnote{A jet is a 3D distribution of energy deposition that is
   usually well contained by a cone (think about a very big elongated
   drop of water, to visualize it).} and physics vectors
 (4-momenta\footnote{A four-momentum is a vector of the spacetime
   whose time-like component is (proportional to) the particle energy
   and the space-like component is the 3D momentum.}).  In addition,
 clusters and physics vectors are used to build tracks that
 visualize the path of particles through the detector.


\section{Description of the {\ROOT} Framework}\label{sec-overview}

 The {\ROOT} framework contains about 3000 classes, grouped into
 about 110 packages and plugins. In addition, the latter are grouped into
 top-level categories that are the subject of this section.


\subsection{Input/Output}\label{sec-io}

\begin{figure}[t!]
  \centering
  \includegraphics[width=0.9\textwidth]{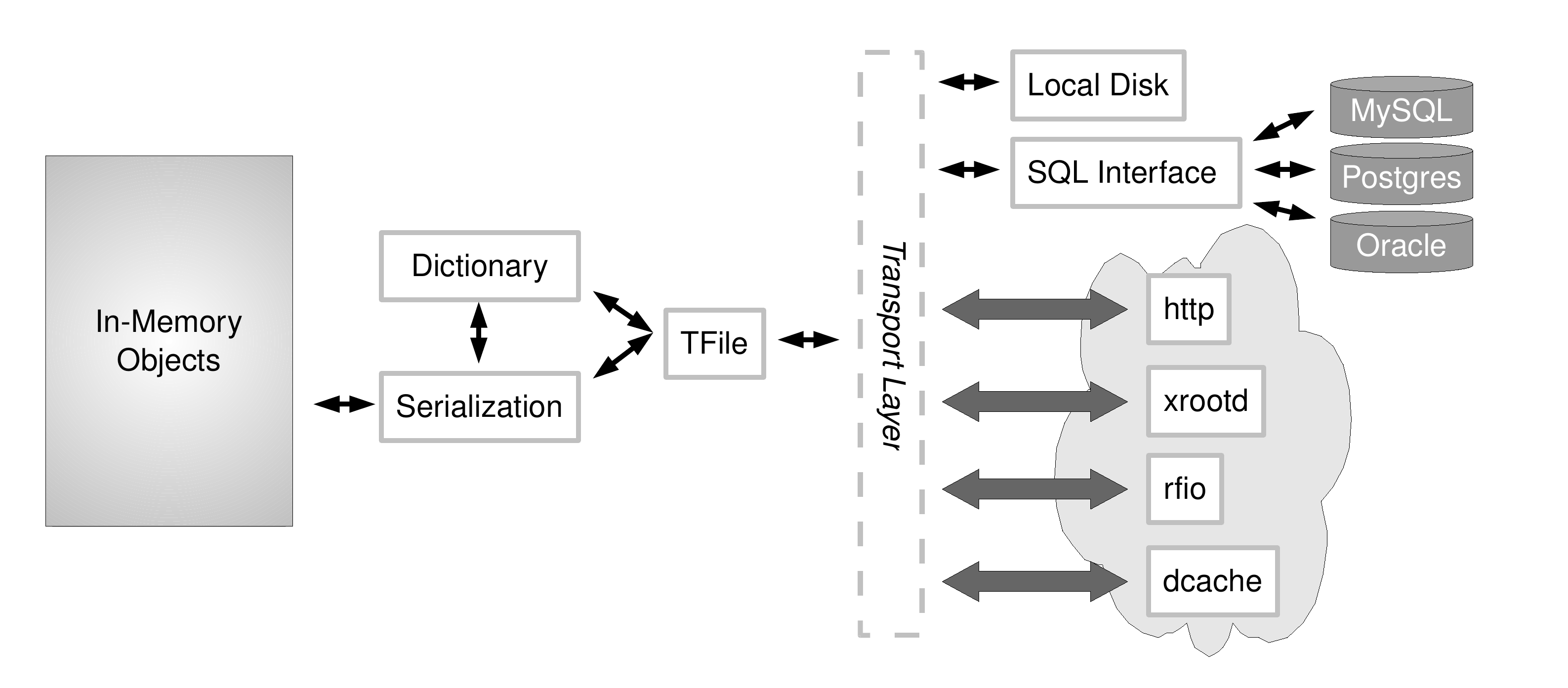}
  \caption{Transient/persistent conversion in {\ROOT}.}\label{fig-io}
\end{figure}

 {\ROOT} is currently used for storing up to 50 petabytes of data according to the latest
 estimates\footnote{According to a survey of a number of experiment computing coordinators}.
 The I/O layer stores C++ objects into storage
 systems, be it file systems, databases, common protocols to storage
 elements (like \Pkg{xrootd} \cite{xrootd},
 \Pkg{dCache}\footnote{\url{http://www.dcache.org/}}, or
 \Pkg{rfio}\footnote{\url{http://hikwww2.fzk.de/hik/orga/ges/infiniband/rfioib.html}}),
 or HTTP, see figure~\ref{fig-io}.


\subsubsection{Describing C++ Objects}\label{sec-dict}

 To be stored, C++ objects need to be described: the I/O must know
 \emph{what} to store.  {\ROOT} provides this description (called
 \emph{dictionary}) for all its classes and users can build dictionaries for their
 own classes.  The description data (commonly called
 \emph{reflection}) are provided by \Cmp{CINT}, or by a combination of
 \Pkg{GCCXML} \cite{gccxml} and \Cmp{Reflex}, a C++ reflection library
 that is part of ROOT.  Based on that information, {\ROOT} knows where in memory
 an object's data members are, what their size is, and how to store
 them.  {\ROOT} I/O supports pointer (un-)swizzling, the conversion of 
 pointers to indexes in the output buffer. 
 It can even deal with an object graph with circular
 references (making sure each object is streamed only once to the
 buffer), and it is able to restore it correctly upon reading.

 Because the description of all relevant classes is stored with the
 data, changes of the class definition of objects stored with {\ROOT}
 I/O are supported.  When reading, the descriptions from the
 persistent layer and the in-memory version are compared: if
 differences are found, {\ROOT} automatically translates in many cases
 from the old to the new format (\emph{schema evolution}).  A complete
 framework for arbitrary user controlled conversions is also available \cite{dme}.


\subsubsection{\Cls{TFile}}\label{sec-tfile}

 A {\ROOT} file is read and written by the class \Cls{TFile} and is designed
 to be write-once, read-many (while supporting deletion and re-use of
 contained data).

 The content of a {\ROOT} file is a simple binary stream, with a
 layout described in the class documentation of \Cls{TFile}
 \cite{ref}.  All data but the header is usually compressed to reduce the
 storage space and I/O bandwidth usage of files at the cost of slightly increased CPU time
 when reading and writing files.  The file consists of a content
 index, the list of type descriptions relevant for the file, and the
 actual data.  Each data chunk is named and it can be retrieved given
 its name.  \Cls{TFile} also supports hierarchical storage in nested
 directories.

 Typical file sizes range from a few kilobytes to several gigabytes.
 Files can be merged into new, larger files; this can be done
 recursively, i.e. merging also the collections themselves that are
 contained in the file, as long as they have the same name and are of
 the same type.  Collections of files can also be merged into a zipped
 container; {\ROOT} supports transparent unzipping of and navigation
 in this collection of files.

 The description of the classes stored in the file (\S\ref{sec-dict})
 can be used to read the data even without the C++ class definition.
 One can thus write C++ objects using the definition from a user
 library, and read them back without the user library.  Any available
 reflection data is used to interactively browse a {\ROOT} file using
 the \Cls{TBrowser} that can also expand and browse the content of
 all C++ objects, either from {\ROOT} or STL, or user defined.

 {\ROOT} files can be opened via the HTTP protocol, without any
 special server requirement.  {\ROOT} only asks for those parts of the file (using
 http \texttt{content-range} requests) that are actually required.  This
 allows a low-latency, live remote browsing of {\ROOT} files.


\subsubsection{\Cls{TTree} and I/O}\label{sec-tree-io}

\begin{figure}[t!]
  \centering
  \includegraphics[width=0.6\textwidth]{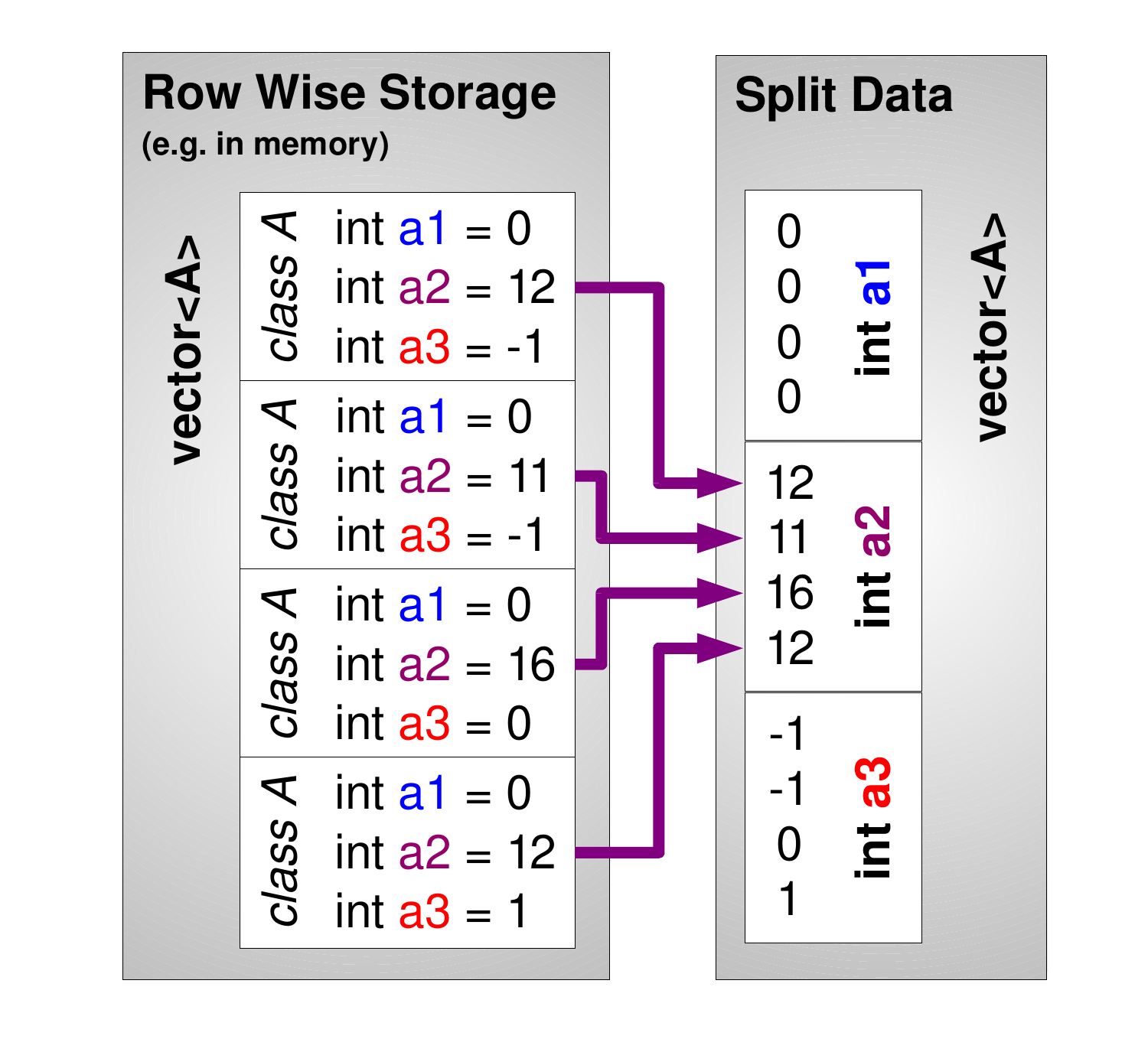}
  \caption{Automatic splitting of a container of objects.}\label{fig-split} 
\end{figure}

 A \Cls{TTree} is a container that is optimized for I/O and memory usage.
 A \Cls{TTree} consists of branches, branches can contain complete objects
 of a given class or be split up into sub-branches containing individual data
 members of the original object. This is called \emph{splitting} and can be done
 recursively till all sub-objects are split into branches only containing individual
 data members. Splitting can even transform containers into branches of the
 containee's data members, grouping them as shown in \ref{fig-split}.
 Splitting can be done automatically using a class' dictionary
 information. Each branch stores its data in one or more
 associated buffers on disk. The desired level of splitting depends on the
 typical future access patterns of a tree.
 If during analysis all data members of an object will be accessed then
 splitting will not be needed. Typical analyses access only a few data
 members; in this case splitting is highly beneficial.

\begin{figure}[t!]
  \centering
  \includegraphics[width=0.9\textwidth]{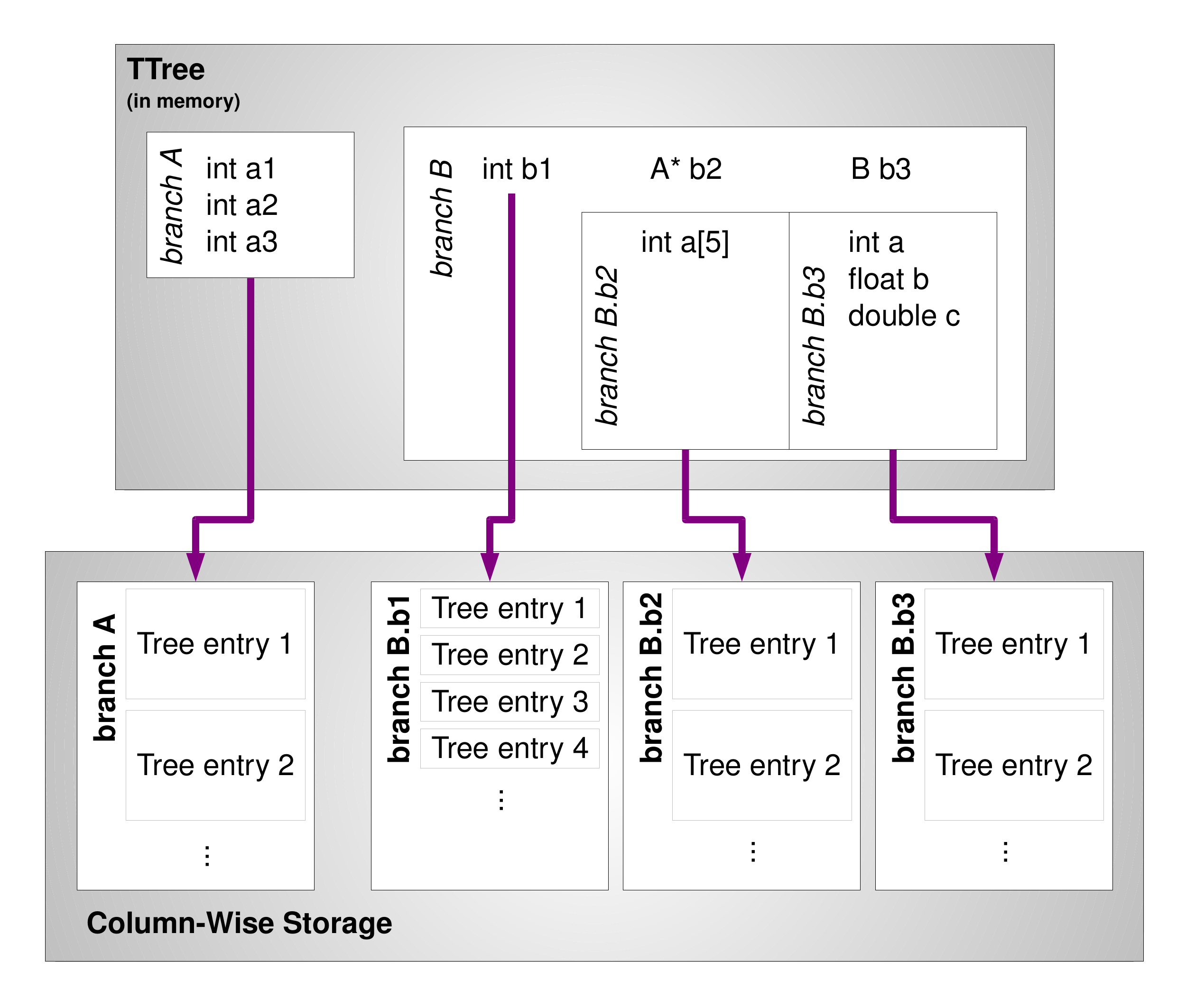}
  \caption{Column-wise layout of \Cls{TTree} data in memory buffers.}\label{fig-cws} 
\end{figure}

 Branch-based storage is called vertical or column-wise storage (CWS;
 figure~\ref{fig-cws}), as opposed to horizontal or row-wise storage (RWS) as is
 usually found in RDBMS databases.  In CWS, just like in RWS, a collection
 (``table'') of similar objects (``rows'') is assumed.  However, in RWS
 all data members of an object are always read, while in CWS only the needed 
 buffers (e.g. data members) are read.
 Splitting is an automated way to create these columns.

 CWS reduces the number of I/O operations and the amount of transferred
 data, because it reads only the needed parts of
 each object.  All other members of the object keep the values defined
 by the class default constructor.  When iterating through the
 collection, data members that need to be read are consecutive on the
 storage medium in the case of CWS.  This allows block-wise reading of
 the data for several entries (rows) in one go, something massively favored
 by all modern operating systems and storage media. Another advantage stems from
 the fact that {\ROOT} compresses the data buffers using Huffman
 encoding \cite{zip}, which benefits from seeing the same byte pattern
 more often, because the same data member usually has similar values
 (e.g.\ a particle's type ID).

 Because a \Cls{TTree} describes the objects it contains, one can read
 objects from a \Cls{TTree} even without their original class
 definition.  The \Cls{TTree} can even generate a C++ header file
 representing the layout of the object's data as stored in the
 \Cls{TTree}.  Combined with the power of the interpreter and {\ACLiC}
 (\S\ref{sec-aclic}) this allows a smooth transition from stored
 binary data to C++ objects, even without C++ libraries.  \Cls{TTree}s
 can also generate a \Cls{TSelector} skeleton (\S\ref{sec-selector})
 for data analysis automatically.

 Given the huge amount of data commonly processed by users of {\ROOT},
 \Cls{TTree}s often do not fit into a single file, or the file grows
 to impractical sizes.  In addition, in (parallel) batch system-based
 analyses, splitting \Cls{TTree}s across several files facilitates the
 distribution of data.  {\ROOT} supports this with \Cls{TChain}, by
 implementing a collection of \Cls{TFile}s that all contain a part of
 the same\footnote{With identical name and struture.} \Cls{TTree}.
 The \Cls{TChain} inherits from \Cls{TTree}, hence making it
 irrelevant to the user whether the \Cls{TTree} is stored in one or
 several physical files.

 Analyses commonly access the same part of a \Cls{TTree} for all its
 entries. {\ROOT} implements an auto-adaptive
 pre-fetch mechanism reading the next entry while the
 previous entry is still being processed.  This
 reduces the effect of high latency networks dramatically:
 reasonable sized analyses become viable even over ADSL. Table~\ref{tab-treecache}
shows the duration of an example data analysis.
 The 280\,MB data file is hosted at CERN with a $100$\,Mbit/sec network
 connection; the analysis reads $6.6$\,MB. The bandwidth shown is the
 smallest bandwidth found on the connection path. For a low-occupancy connection
 bandwidth is clearly not the limiting factor.

\begin{figure}
\center
{\scriptsize
\vspace{1em}
\begin{tabular}{|l|r|r|l|r r r|}
\hline
\multirow{2}{*}{\parbox[b]{7em}{\bf Location of \\ Data Analysis}} 
  & \multirow{2}{*}{\parbox[b]{5.5em}{\bf Bandwidth \\ (Mbit/s)}}
  & \multirow{2}{4em}{\parbox[b]{3em}{\bf Latency \\ (ms)}}
  & {\bf Analysis CPU}
  & \multicolumn{3}{c|}{\bf Duration (s)} \\
\cline{5-7}
& & & & \multicolumn{3}{c|}{\bf Cache size (KB)} \\
& & & & {\bf 0} & {\bf 64} & {\bf 10240} \\
\hline
\hline
local (no network)  &  &  & Pentium4, 2.4\,GHz   & 3.4 & 3.4 & 3.4 \\
CERN  & 100 & 0.3 & Pentium4, 3\,GHz & 8.0 & 6.0 & 4.0 \\
CERN wireless & 10  & 2.0 & Core Duo, 2\,GHz & 12 & 5.6 & 4.9 \\
Orsay, France         & 100  & 11.0 & Pentium4, 3\,GHz& 130 & 12 & 9.0 \\
Amsterdam, NL & 100  & 22.0 & Opteron 280 & 230 & 12 & 8.4 \\
ADSL & 8  & 72.0 & Core Duo, 2\,GHz & 740 & 48 & 28 \\
Caltech, USA & 10,000  & 240.0 & Opteron 280 & $>1,800$ & 130 & 9.9 \\
\hline
\end{tabular}
\vspace{1em}
\caption{Performance improvements by the \Cls{TTree} cache, see text.}
\label{tab-treecache}
}
\end{figure}


\subsubsection{I/O Formats}

 {\ROOT} can store via its I/O interface any C++ objects
 in binary {\ROOT} files.  It also supports the XML representation,
 though mostly for didactic purposes\footnote{This format has been
   implemented originally as an exchange format with non ROOT based
   applications, but only a few applications have made use of it.}: it
 nicely demonstrates the layout, but its performance (due to XML's
 ASCII-based representation) and disk usage (due to XML's verbose
 meta-data) prohibits its used as a production storage format.

 Data can also be stored into database tables through an abstraction
 layer; the description of objects and their member is translated into
 tables and their columns.


\subsection{Mathematical and Statistical Tools}\label{sec-math-stat}

 One may need to manipulate data in a number of different ways.
 Because {\ROOT} is a C++ framework, all C and C++ standard functions
 are available.  In addition, {\ROOT} provides a number of advanced
 mathematical and statistical functions, well integrated into the
 framework, that allow to perform virtually all possible operations
 with a few simple commands.

 The minimal set of tools required for numerical computing is provided
 by the \Cmp{MathCore} library.  It consists of the following
 components.
\begin{itemize}
  \item Commonly used mathematical functions like special functions
        not provided yet by the C++ standard and statistical
        distribution functions. For each statistical distribution, the
        probability density, the cumulative and its inverse functions
        are provided.  These functions are provided 
        in the namespaces \NmSp{ROOT::Math} and \NmSp{TMath}.

  \item Classes for random number generations (\Cls{TRandom}
        classes). The default pseudo-random number generator is the
        Mersenne and Twister generator (\Cls{TRandom3} class)
        \cite{MersenneTwister}.

  \item Basic implementation and interfaces of numerical algorithms,
        like integration, derivation or simple (one dimensional)
        minimization.

  \item Classes and interfaces required for fitting all the {\ROOT}
        data objects.

  \item Abstract interfaces and adapter classes for function
        evaluation in one or more dimensions.
\end{itemize}

 The \Cmp{MathMore} library complements \Cmp{MathCore} by providing
 additional mathematical functionality. It is based on the GNU
 Scientific Library (\Pkg{GSL}) \cite{gsl}, which is used as an
 external library.  \Cmp{MathMore} implements extra special
 functions like Bessel functions of various types and fractional
 order, elliptic integrals, Laguerre and Legendre polynomials,
 hypergeometric functions.  \Cmp{MathMore} contains additional
 implementations of the numerical algorithms and extra random number
 generators which are present in \Pkg{GSL}.

 Various libraries exist for numerical minimization and fitting.
 These libraries include the numerical methods for solving the fitting
 problem by finding minimum of multi-dimensional functions.  A common
 interface exists in \Cmp{MathCore} (the class
 \NmSp{ROOT::Math::}\Cls{Minimizer}) for multi-dimensional numerical
 minimization.  Several implementations of this interface are
 present in {\ROOT}:
\begin{itemize}
  \item \Cmp{Minuit} provides an implementation of the popular
        \Pkg{MINUIT} minimization package \cite{minuit}.  It is a direct
        translation from the original Fortran code to C++ and provides
        a very similar API.

  \item \Cmp{Minuit2} is a completely new objected-oriented
        implementation of \Pkg{MINUIT} \cite{minuit2}.  The same
        minimization algorithms like \Alg{Migrad} and \Alg{Simplex}
        are present, but with new objected-oriented interfaces.  
        Furthermore, it provides an implementation of a specialized method for
        finding the minimum of a standard least-square or likelihood
        functions, by linearizing the Hessian matrix. This algorithm is called in ROOT 
        \Alg{Fumili2}.

  \item \Cmp{Fumili:} library providing the implementation of the \Alg{Fumili} fitting algorithm\cite{fumili}, 
        another specialized minimization method for least-square or likelihood functions.

  \item \Cmp{MathMore} offers minimizers based on \Pkg{GSL}. 
        These include various minimization methods
        based on conjugate gradient algorithms, the
        Levenberg-Marquardt algorithm \cite{lma} for non-linear
        least-squares fitting and a stochastic minimization method
        based on simulated annealing.

  \item The \Cls{TLinearFitter} class implements linear least-squares
        fitting with a possibility for using robust fitting.
\end{itemize}

{\ROOT} contains two libraries providing matrices and vector classes
 and linear algebra operations:
\begin{itemize}
  \item \Cmp{Matrix}: general matrix package including matrix
        \Cls{TMatrix} and vector \Cls{TVector} classes and the
        complete environment to perform linear algebra calculations,
        like equation solving and eigenvalue decompositions.

  \item \Cmp{SMatrix}: package optimized for high performance matrix
        and vector computations of small and fixed size.  It is based
        on expression templates to achieve a high level optimization
        and to minimize memory allocation in matrix operations.  It
        derives from a package originally developed for HeraB
        \cite{smatrix}. Performance studies of the matrix packages in
        benchmark applications used in HEP have been shown elsewhere
        \cite{mathChep07}.
\end{itemize}

 Two libraries exist in {\ROOT} also for describing physics vectors in 2,
 3 and 4 dimensions (relativistic vectors) with rotation and
 transformation algorithms: 
\begin{itemize}
  \item \Cmp{Physics}: library with the \Cls{TVector3} and
        \Cls{TLorentzVector} classes.

  \item \Cmp{GenVector}: package with generic class templates for
        modeling geometric vectors in 2 and 3 dimensions and Lorentz
        vectors.  The user may control how the vector is internally
        represented, by making a choice of coordinate system and
        underlying scalar type.
\end{itemize}

 Other mathematical and statistical packages in {\ROOT} are:
\begin{itemize}
  \item \Cmp{Unuran}: universal algorithms for generating non-uniform
        pseudo-random numbers, from a large set of classes of
        continuous or discrete distributions in one or several
        dimensions\footnote{\url{http://statmath.wu-wien.ac.at/unuran/}}.

  \item \Cmp{Foam}: multi-dimensional general purpose Monte Carlo
        event generator (and integrator). It randomly generates points
        (vectors) according to an arbitrary probability distribution
        in $n$ dimensions.\cite{foam}

  \item \Cmp{FFTW}: library with implementation of the fast Fourier
        transform (FFT) using the \Pkg{FFTW} package\footnote{The
        ``Fastest Fourier Transform in the West'',
        \url{http://www.fftw.org/}}. It requires a previous 
        installation of \Pkg{FFTW}.

  \item \Cmp{MLP}: library with the neural network class,
        \Cls{TMultiLayerPerceptron} based on the \Alg{NN} algorithm from the
        \Pkg{mlpfit}
        package\footnote{\url{http://schwind.web.cern.ch/schwind/MLPfit.html}}.

  \item \Cmp{Quadp}: optimization library with linear and quadratic
        programming methods. It is based on the \Pkg{Matrix} package.

  \item \Cmp{Statistic classes for computing limits and confidence
        levels}. Some of these classes are currently provided by
        \Lib{libPhysics}.

  \item \Cmp{TMVA}: toolkit for multivariate data analysis, providing
        machine learning environment for the processing and parallel
        evaluation of sophisticated multivariate classification
        techniques. Though specifically designed to the needs of
        high-energy physics applications, it offers general methods
        that can be used in other fields, too \cite{tmva}.

  \item \Cmp{RooFit}: toolkit for modeling statistical distributions
        (especially the ones used in physics analysis). Models can be
        used to perform likelihood fits, produce plots, and generate
        ``toy Monte Carlo'' samples for various studies \cite{roofit}.

  \item \Cmp{RooStats}: package providing the required advanced
        statistical tools needed by the LHC experiments for their
        final data analysis in order to calculate confidence
        intervals, to perform hypothesis tests and combinations of
        different analysis channels.  It provides common interfaces to
        the major tools with implementations based on different
        statistical techniques, which have been approved by the
        experiment statistical committees. It is based on the
        \Cmp{RooFit} classes for describing probability density
        functions or likelihood functions.
\end{itemize}


\subsection{Histograms}\label{sec-histo}

 When dealing with many events, one usually adopts statistical methods
 to analyze them. Two different approaches are possible:
 statistical data analysis of binned or unbinned data.  The most
 frequently used approach involves binned data, in the form of
 histograms, whereas unbinned data are saved into instances of the
 \Cls{TTree} class (see \S\ref{sec-tree-io}).

 In {\ROOT}, 1-dimensional histograms are defined by the base class
 \Cls{TH1}: actual classes inherit from \Cls{TH1} with the type of 
 the bin count (char, float, double,...) defined by the derived class.
 \Cls{TH1} is also the base class for 2D
 and 3D histograms (again, supporting different types of entries) and
 for profile histograms (\Cls{TProfile}, \Cls{TProfile2D} and
 \Cls{TProfile3D}).  Profile histograms are used to display the mean
 value of a variable and its standard deviation in each bin of
 another dependent variable (or variables in case of multi-dimensional
 profile histograms). Histogram classes can also be used to analyze
 weighted data sets.

 {\ROOT} histograms internally contain a pair (value, uncertainty) for
 each bin, plus the numbers of entries which fall outside its limits
 (both overflow and underflow).  Additional information like the
 total number of entries and the integral of the histogram are also
 stored.  Statistical information such as the mean and standard
 deviation along the histogram axis can be obtained.  The binning can
 be defined with constant or variable step size and higher-dimensional
 histograms support projecting and slicing.  Histograms can also be
 fitted with a user provided function.

 Many types of operations are supported on histograms or between histograms:
addition and subtraction, multiplication and division with histograms, 
functions, or scalars. They can also be rebinned and compared using statistical
 hypothesis tests like the chi-square test.

 Histograms can be plotted by invoking the \Mtd{Draw()} method and the
 result can be interactively manipulated (see \S\ref{sec-graphics}).
 Labels can be numerical or textual and the user can define
 titles\footnote{\LaTeX-like strings are supported.} for the
 histogram and each axis.

 Sets of $(x,y)$ or $(x,y,z)$ data can be displayed and analyzed in
 {\ROOT} using the \Cls{TGraph} or \Cls{TGraph2D} classes. The data
 errors can also be displayed using the derived classes
 \Cls{TGraphErrors} and \Cls{TGraphAsymErrors}.  In addition to
 fitting, the \Cls{TGraph} classes provide the functionality for
 interpolating the data points using different techniques such as
 cubic splines and for smoothing.

 {\ROOT} allows the user to fit both binned and unbinned
 data with parametric functions which can be displayed together with the 
data. The plottable functions are represented by the classes \Cls{TF1},\Cls{TF2}
or \Cls{TF3} depending on the dimension. They can be created either from  precompiled user code, 
using global functions or class member functions or from mathematical expressions which are  handled 
by the \Cls{TFormula} class.  \Cls{TFormula} is able to parse expressions containing  mathematical functions, 
including those in \Cls{TMath}  and using a  special syntax for defining the parameters.  
Predefined expression representing functions like polynomial, Gaussians, exponential or Landau
are also available to facilitate the usage.   

\begin{figure}[t!]
  \centering
  \includegraphics[width=\textwidth]{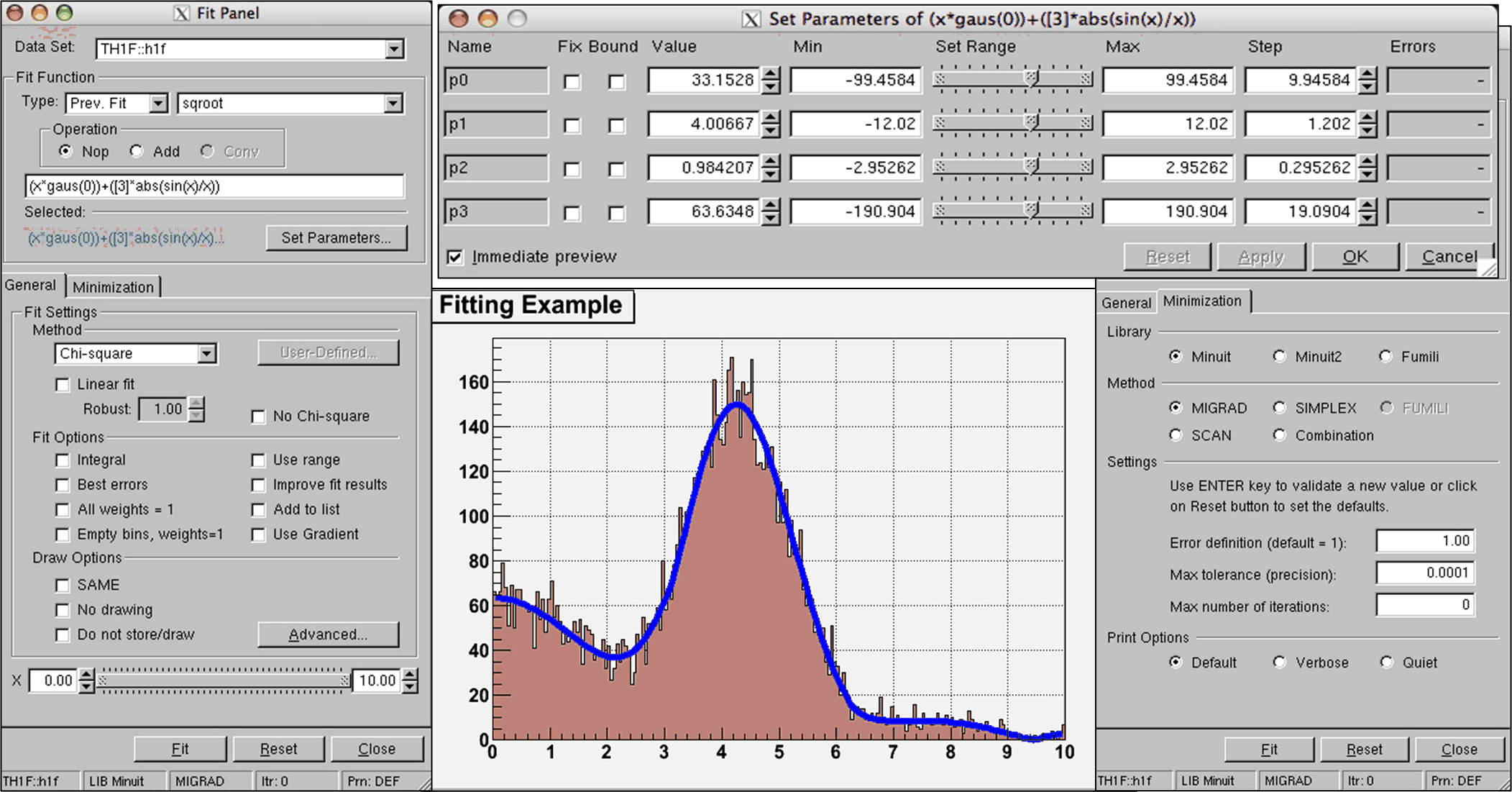}
  \caption{The ROOT fit panel: the General tab (left) for selecting
           function, fit methods and options, the Set Parameter dialog
           (up right) for setting initial values and limits, and the
           Minimization tab (bottom right) for selecting the
           minimization library and method.}\label{fig-fitpanel}
\end{figure}

In addition to invoking the \Mtd{Fit()} method from a macro, the
 user can also make use of the GUI provided by the fit panel
 (figure~\ref{fig-fitpanel}) during interactive sessions.  It can be
 opened directly from the {\ROOT} \Cls{TCanvas} menu or via the
 context menu of any {\ROOT} object which is suitable for fitting,
 available after a right mouse click on the object.  With the fit
 panel, the user can select the fit function, set the initial
 parameter and control all the available fit options.  It offers also
 the possibility to draw scan plots and contour plots of the fitted
 parameters.

\subsection{Graphics and User Interface}\label{sec-graphics}

 Whenever {\ROOT} draws an object, it puts it into a \Cls{TCanvas}
 instance, representing an area mapped to a window directly under the
 control of the display manager.  One can save the \Cls{TCanvas} into
 several possible formats:
 for standard graphics formats, publication quality is obtained by means of
 vector graphics like PostScript or PDF,
 but raster graphics is usually a better choice for images to be
 included into web pages. One can also store it as a C++ macro where 
 the C++ statements reproduce the state of the \Cls{TCanvas} and its contents.
 This allows complete reproduction from within {\ROOT}.

\begin{figure}[t!]
  \centering
  \includegraphics[width=0.9\textwidth]{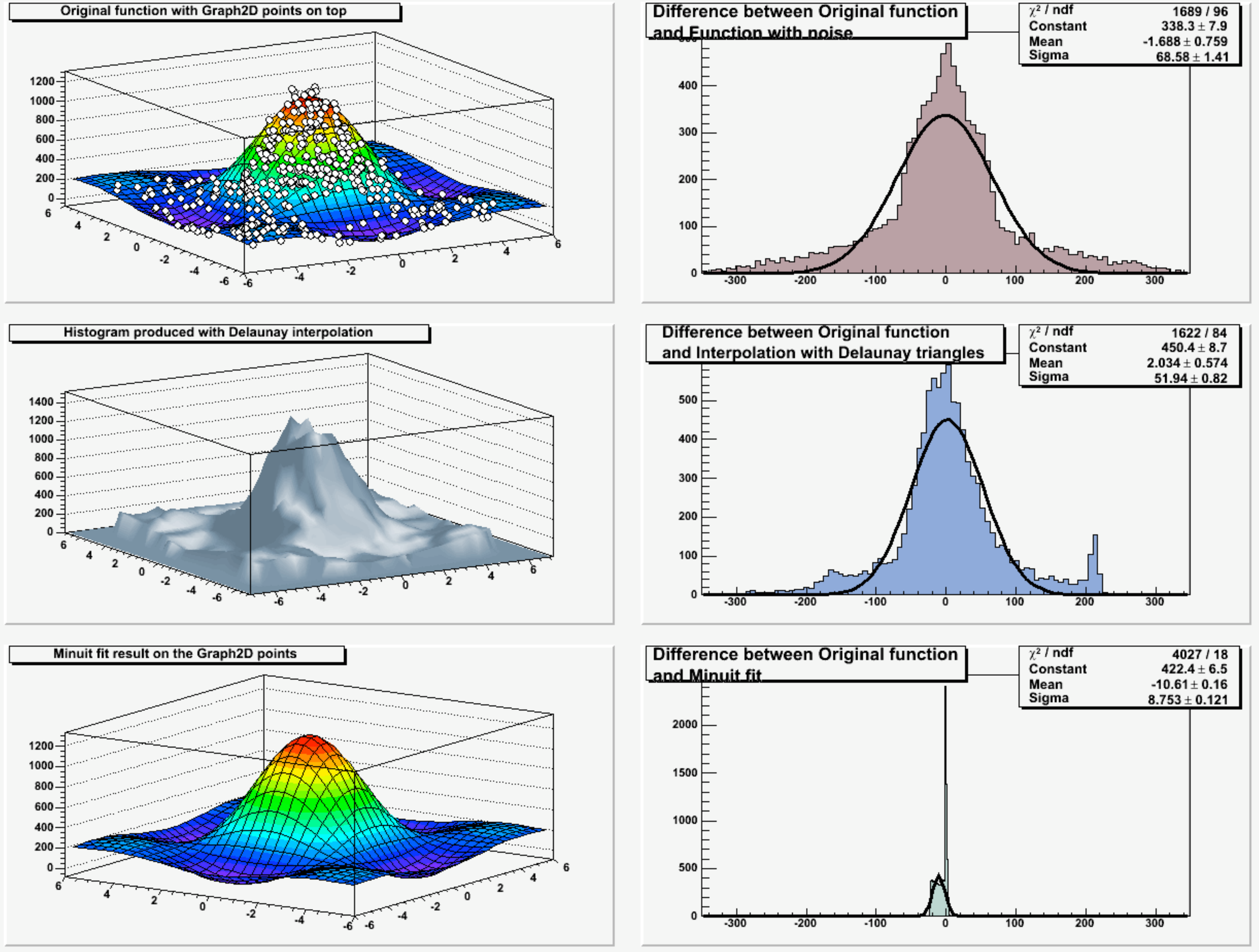}
  \caption{Example of graphical output.  The canvas contains 6
           pads.}\label{fig-graph-example} 
\end{figure}

 Of course, we can open multiple canvases if we want to display
 different things, but it is often better to organize everything into
 a single \Cls{TCanvas}. For this reason, a \Cls{TCanvas} instance can
 be subdivided into independent graphical areas, called ``pads'' (by
 default, a canvas contains a single pad, occupying the whole space ---
 \Cls{TCanvas} inherits from \Cls{TPad}), as shown in
 figure~\ref{fig-graph-example}.

 All {\ROOT} classes inheriting from \Cls{TObject} can be displayed
 on a pad with the \Mtd{Draw()} method.  Graphical object sizes are
 usually expressed in user coordinates.  For instance, after a
 histogram or a graph has been drawn, the user coordinates coincide
 with those defined by the plot axes. The pad position in its parent
 pad is expressed in normalized coordinates,
 in which the pad is mapped to a unit rectangle.
 The \Cls{TCanvas} requires dimensions in pixels to be positioned on
 the desktop.

 In {\ROOT}, the \Mtd{Draw()} method does not actually draw the object
 itself. Rather, it adds the object to the display list of the pad
 (so that it gets drawn every time the pad is redrawn) and invokes the
 \Mtd{Paint()} method, that draws the actual graphics primitives.
 {\ROOT} manages the repainting of the \Cls{TCanvas} automatically
 when either the object is updated of the operating system requires.

 Every {\ROOT} object drawn on a pad can be edited interactively. In
 addition to the pop-up editor (opened from the menu obtained by
 right-clicking on any object), each canvas can also host an editor
 (opened by selecting ``Editor'' from the ``View'' menu provided by the
 window). To modify any object shown by the canvas, simply open the
 latter editor and click on the object.


\subsubsection{2D Graphics}

 2D graphics include everything we can display on the monitor or print
 on paper. {\ROOT} needs to be interfaced with the operating system's
 graphics engine, in order to be able to display windows containing
 some plot, for example. {\ROOT} uses the \Pkg{X11} graphics engine on
 unix-like systems and \Pkg{Win32} on Windows, but can also use the
 multi-platform \Pkg{Qt} library\footnote{Originally provided by
   Trolltech, that was renamed to Qt Software
   (\url{http://www.qtsoftware.com/}) after acquisition by Nokia in
   2008.}.

 Through the \Lib{libAfterImage}
 library\footnote{\url{http://www.afterstep.org/afterimage/}}, {\ROOT}
 is also able to load bitmap images and to manipulate them.  This
 package also allows to produce bitmap output files in all common
 formats such as GIF, PNG, JPEG, etc.


\subsubsection{3D Graphics}

 There are several ways to render 3D graphics in {\ROOT}, the
 preferred one using the
 \Pkg{OpenGL}\footnote{\url{http://www.opengl.org/}} graphics library,
 which is used in {\ROOT} to display data using lego and surface plots
 and to render detector geometries.  Work is in progress to also use
 it for 2D graphics and thus have a single, portable rendering
 interface for 2D and 3D screen graphics.


\subsubsection{Geometry and Event Display}

Geometry in the 3D space is described in {\ROOT} by means of basic
 solids that can be joined, intersected or subtracted to create more
 complex shapes. The possibility to visualize 3D objects is very
 important.  {\ROOT} implements its own scene-graph management library
 and rendering engine that provides advanced visualization features
 and real-time animations. \Pkg{OpenGL} library is used for actual
 rendering.

\begin{figure}[t!]
  \centering
  \includegraphics[width=\textwidth]{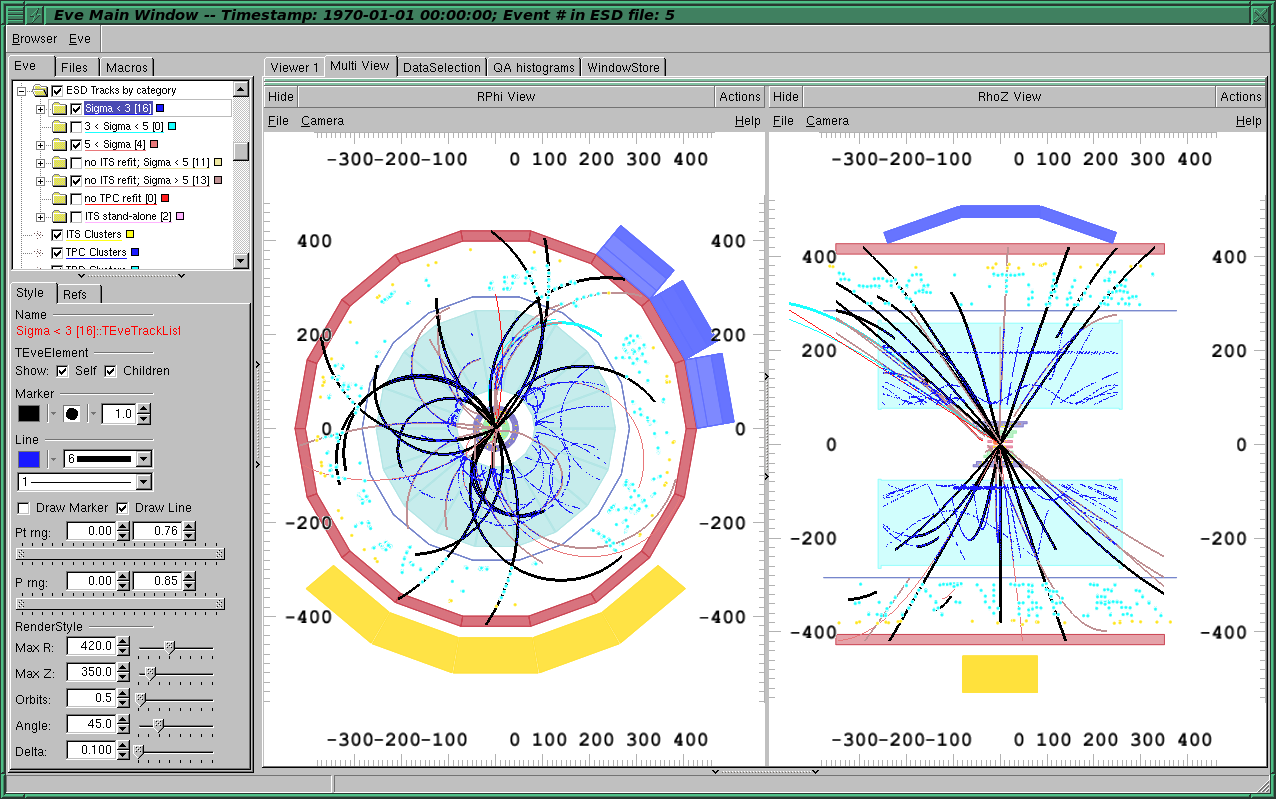}
  \caption{Screenshot of \texttt{AliEVE} showing a simulated
           proton-proton event at the LHC collider as seen by the
           ALICE detector. The reconstructed particle trajectories are
           shown as black lines and the measured particle
           passage-points as colored dots.}\label{fig-alieve}
\end{figure}

 Event display programs are an important application of 3D
 visualization.  \Pkg{EVE}, the event visualization environment of
 {\ROOT}, uses extensively its data-processing, GUI and \Pkg{OpenGL}
 interfaces. \Pkg{EVE} can serve as a framework for object management
 offering hierarchical data organization, object interaction and
 visualization via GUI and \Pkg{OpenGL} representations and automatic
 creation of 2D projected views. On the other hand, it can serve as a
 toolkit satisfying most HEP requirements, allowing visualization of
 geometry, simulated and reconstructed data such as hits, clusters,
 tracks and calorimeter information. Special classes are available for
 visualization of raw-data and detector response. \Pkg{EVE} is used in
 the
 ALICE\footnote{\url{http://aliceinfo.cern.ch/Public/Welcome.html}}
 experiment as the standard visualization tool, \texttt{AliEVE}
 (figure~\ref{fig-alieve}), using the full feature set of the
 environment. In the
 CMS\footnote{\url{http://cms.web.cern.ch/cms/index.html}} experiment,
 \Pkg{EVE} is used as the underlying toolkit of the \texttt{cmsShow}
 physics-analysis oriented event-display. Both \texttt{AliEVE} and
 \texttt{cmsShow} are also used for the online data-quality
 monitoring.

\subsubsection{Graphical User Interface}

 The {\ROOT} Graphical User Interface (GUI) integrates typical GUI
 functionality with {\ROOT} features, like storing the GUI as C++ source, interpreted
 GUI via \Cmp{CINT} and \Cmp{CINT}-based signal/slot
 communication. The result is a flexible GUI toolkit, rich of
 functionalities and offering all widgets that are provided by other
 toolkits, including a GUI builder\footnote{The development of a
   dedicated {\ROOT} GUI was required because when the project started
   there were no good cross platform toolkit; \Pkg{Qt} existed but had
   license problems.}.

 The {\ROOT} \Cmp{GUI builder} provides tools for developing user
 interfaces based on the {\ROOT} GUI classes. It offers a palette of
 user interface elements. They can be selected, positioned, and
 grouped, laid out in the main application frame. According to the
 selected widget, a dynamically created context menu provides detailed
 control of widget attribute settings. One can save on a {\ROOT} macro
 the result, and take such C++ code as starting point for further
 developments.


\subsection{Simulation}\label{sec-simulation}


 \Cls{TVirtualMC} provides a virtual interface to Monte Carlo
 applications, allowing the user to build a simulation independent of
 any actual underlying Monte Carlo implementation itself. A user will
 have to implement a class derived from the abstract Monte Carlo
 application class, and provide functions like
 \Mtd{ConstructGeometry()}, \Mtd{BeginEvent()}, \Mtd{FinishEvent()},
 \ldots The concrete Monte Carlo implementation (\Pkg{Geant3},
 \Pkg{Geant4}, \Pkg{Fluka}) is selected at run time --- when
 processing a {\ROOT} macro where the concrete Monte Carlo object is
 instantiated. This allows for comparison between different engines
 (often used to estimate the systematic simulation uncertainties) using a single
 application.
 {\ROOT} thus offers a single interface common to all of the most common simulation
 engines; it offers a centrally managed, performant C++ geometry system instead
 of a plethora of different, often incompatible and too specialized geometry systems
 as provided by the simulation engines. Its geometry system offers I/O capabilities
 and an interface to {\ROOT}'s event display.
 Examples of \Cmp{VMC} can be found in AliROOT\cite{aliroot} for the
 ALICE experiment at the LHC or FAIRROOT\cite{fairroot} for the
 FAIR experiments at GSI, Darmstadt.

 Monte Carlo simulations always have to describe the input particles,
 together with their interactions, and the detector (geometry,
 materials and read-out electronics). The definition of particles,
 available interactions and detector is done during the
 initialization phase. The main body of the application is then a loop
 over all particles that are traced through all materials until they
 exit, stop or disappear (by decay or annihilation). The tracing is
 done in a discrete fashion: at each step, the detector volume is
 found in which the particle is located and pseudo-random numbers are
 used to ``draw'' one among possibly several physical processes, to
 simulate the interaction of the particle with the matter. If an
 interaction occurs, the energy lost by the particle is computed
 (again, it is usually a random process) and subtracted from its
 kinetic energy. When the latter reaches zero, the particle stops,
 otherwise a new step is performed. 

 Having computed the energy lost by all particles inside the detector,
 one has to simulate the behavior of the read-out electronics. This is
 usually done later, with another program that receives the energy
 lost in different locations as input, but it can also be done by the
 very same application that is performing the particle tracing inside
 the detector. Usually, the simulation of the read-out electronics
 also involves some use of pseudo-random generators, at least to
 simulate the finite resolution of any real measuring device.

 In any detector simulation, the definition of its geometry has
 special importance.  The {\ROOT} \Cmp{geometry package} is a tool to
 build, browse and visualize detector geometries. It is independent
 from any Monte Carlo simulation engine, though it has been designed to
 optimize particle transport in correlation with simulation packages
 as \Pkg{Geant3}, \Pkg{Geant4} and \Pkg{Fluka}.

 Most detectors in HEP have been modelled with the {\ROOT} geometry
 (experiments at LEP, LHC, FNAL, HERA, GSI, etc.).
 For example, the standard {\ROOT}
 test suite tracks particles to 35 large detectors.  The Geometry
 Description Markup Language
 (GDML)\footnote{\url{http://gdml.web.cern.ch/GDML/}} system can be
 used to export/import geometries from/to other formats (e.g.\
 \Pkg{Geant3}, \Pkg{Geant4}).

 The building blocks of any geometry are the volumes. Volumes may contain
 other volumes, producing a hierarchy of volumes. The biggest one, 
 called the ``world'', contains all
 other volumes and provides the master reference system (MARS) in
 which the others are positioned. Each volume (except for the "world")
 needs to be associated
 with a medium, that can be a mixture of different materials (whose
 weights are the relative densities).  

 Complex geometries can be built in a hierarchical way, through the
 concept of containment: one has to define and position some volumes
 inside other ones. Positioning is done with spatial transformations
 with respect to the ``mother reference system'' (i.e.\ the system
 defined by the containing volume). Complex volumes are built using
 basic or primitive shapes, already defined by {\ROOT} (e.g.\ box, tube,
 cone, etc.), through operations like join or subtract. Finally, a
 given volume can be positioned several times in the geometry or it
 can be divided accordingly to user-defined patterns, automatically
 defining new contained volumes. 

 Once a geometry has been created, it can be saved into a {\ROOT} file or
 as C++ macro with the \Mtd{Export()} method of
 \Cls{TGeoManager}. Loading the geometry is done with its
 \Mtd{Import()} method. In addition, individual volumes can also be
 saved into a {\ROOT} file. Finally, {\ROOT} provides a graphical user
 interface to edit or build a geometry. The editor can be opened with
 the \Mtd{Edit()} method of \Cls{TGeoManager}.

 Having defined the detector geometry, particles need to be tracked
 inside all volumes, and their interaction simulated.  The application
 can make use of the {\ROOT} \Cmp{geometry package} to build a detector
 and the virtual Monte Carlo interface to access one or more
 simulation engines. {\ROOT} makes it possible also to store and visualize tracks,
 as it is done inside the drawing package with the \Cls{TGeoTrack}
 class.


\subsection{Interpreters}\label{sec-bindings}

 \Cmp{CINT} is an almost full ANSI compliant C/C++ interpreter.  It
 serves as {\ROOT}'s non-graphical user interface, both for
 interactive use (through \Cmp{CINT}'s prompt) and in headless
 ``batch'' mode, where \Cmp{CINT} processes C++ code without showing
 any graphics.  Other use cases are shown in \S\ref{sec-interpuse}.

 In most cases, physicists develop data analysis programs gradually,
 through repeated cycles of changing and running the code.
 Traditionally, the code needed to be compiled, linked, loaded, and
 then again unloaded so the next iteration could be started.  The
 ability to use an interpreter is a fundamental improvement for this
 approach of rapid development.

 \Cmp{CINT} allows interpreted and compiled code to interact: it can
 call compiled code just like it can be called from compiled code, in
 a re-entrant way.  With that, code like \texttt{histogram->Draw()} can be
 interpreted, resulting in the function \texttt{TH1::Draw()} in one of {\ROOT}'s libraries 
 being called. On the other hand, compiled code can contain the statement
 \texttt{gROOT->ProcessLine("myobj->Go()")}, which could execute the interpreted
 function \texttt{MyObj::Go()}.
 The transition of the call chain from interpreted
 to compiled code happens through stubs; \Cmp{CINT} keeps a function
 pointer to the stub for each function that can be called from the
 interpreter.  The stubs are generated as part of the dictionary, see
 \S\ref{sec-dict}.

 {\ROOT} also provides the Python interface \Cmp{PyROOT}\cite{pyroot} that uses some of \Cmp{CINT} features.
 This allows it to do dynamic call translation
 instead of relying on a fixed wrapper.  Also provided is an interface
 to Ruby.
 Python and Ruby offer late binding and an easy to learn syntax.
 For a C++ framework, the major advantage of providing a C++
 interpreter (e.g. compared with a Python interpreter) is the
 homogeneity of languages: users write compiled and interpreted code
 in the same language, they can transfer code or parts of it from the
 interpreted ``world'' to the compiled one without any transition.


\subsubsection{Interpreter Use Cases}\label{sec-interpuse}

 While interpreters' use cases are virtually unlimited, there are
 several key examples of use already in {\ROOT}'s context.  The
 graphical user interface implements the signal slot mechanism through
 the interpreter: the signal is emitted as strings interpreted by the
 interpreter, which are evaluated dynamically.
 This allows powerful expressions and loose coupling
 between the caller and the callee, because the called function does
 not need to be resolved at link time.

 Another use case is {\ROOT}'s auto-documentation component: it parses
 sources on demand, extracting documentation strings.  It can even
 interpret code that is embedded in the documentation, run it, and
 embed the output and the code into the documentation.  This is an
 elegant way of keeping graphical output up to date and of showing
 examples of use for the documented class.

 As already mentioned for signal/slot, the interpreter allows a loose
 coupling of libraries through strings resolved at runtime, instead of
 symbols resolved at link time.  {\ROOT} makes use of this feature for
 its plugin manager: instead of hard-wiring dependencies or
 implementations of interfaces at link time, {\ROOT} determines the
 plugin to use at run time, by executing a registered piece of C++
 code that will instantiate the plugin. This approach
 is dynamic and extensible, eeven by the user.
 It saves resources because it does not load unused plugins.

 {\ROOT} even relies on \Cmp{CINT} for some parts of the I/O
 framework: the interpreter allows {\ROOT} to call a helper function
 on an object given only its memory address and type name.
 This, too, is an ideal use case for an interpreter.


\subsubsection{Automatic Library Builds}\label{sec-aclic}

 Interpreting code is always slower than compiled code.  Once
 code has been developed it should thus be ``moved into the compiled
 world'' and the transitioning of code should be seamless.  But it is
 not: code needs to be compiled, linked, and loaded.  {\ROOT}'s
 serialization framework and the interpreter require an additional
 build step, see \S\ref{sec-dict}. For that, the interpreter
 scans the user's header files and generates a source file containing the
 dictionary. These dictionaries, too, need to be compiled, linked, and loaded.

 Tracking of dependencies is a common
 request, to only update the binary if a relevant source
 file has been changed.
 Traditionally, users would write a Makefile to compile the code which
 they then link into a binary, either into a shared library to be loaded
 into ROOT, or into a stand-alone executable.
 This is a symptom that the migration of code from the interpreter
 to a binary is far from smooth.

 {\ROOT} removes this hurdle altogether, by completely hiding the
 complexity from the user.  To load the source file
 \texttt{myCode.cxx} into the interpreter, one would usually call
\begin{quote}
\begin{verbatim}
.L myCode.cxx
\end{verbatim}
\end{quote}
 This file's functions and types are then available for interpretation.

 To instead load the file as a shared library, and if needed to build
 it on the fly, users issue this command:
\begin{quote}
\begin{verbatim}
.L myCode.cxx+
\end{verbatim}
\end{quote}
 This invokes an integrated build system called {\ACLiC} that
 works on all supported platforms.  It is a powerful replacement for
 external build systems hiding all of the build complexity. 
 Multiple source files can be compiled into a library by including them in a
 wrapper source file.

 The smooth transition from interpreted to compiled code offered by {\ACLiC}
 has been so successful that {\ROOT} is now considering
 the implementation of true just-in-time compilation made possible
 e.g.\ though LLVM \cite{llvm}, \cite{interp}, instead of the invocation
 of external tools through {\ACLiC}.


\subsection{Parallel Processing Using PROOF}\label{sec-parallel}

 The Parallel {\ROOT} Facility, \Cmp{PROOF} \cite{proof}, is an
 extension of {\ROOT} enabling interactive analysis of large sets of
 {\ROOT} files in parallel on clusters of computers or many-core
 machines. More generally \Cmp{PROOF} can parallelize the class of
 tasks for which solutions can be formulated as a set of independent
 sub-tasks ({\em embarrassingly} or {\em ideally} parallel).

 The main design goals for the \Cmp{PROOF} system are:

\begin{itemize}
\item Transparency: there should be as little difference as possible
      between a local {\ROOT} based analysis session and a remote
      parallel \Cmp{PROOF} session. Typically analysis macros should
      work unchanged.
\item Scalability: the basic architecture should not put any implicit
      limitations on the number of computers that can be used in
      parallel.
\item Adaptability: the system should be able to adapt itself to
      variations in the remote environment (changing load on the
      cluster nodes, network interruptions, etc.).
\end{itemize}

 \Cmp{PROOF} is primarily meant as an alternative to batch systems for
 Central Analysis Facilities and departmental work groups (Tier-2's
 and Tier-3's \cite{monarc}) in particle physics experiments. However, thanks to a
 multi-tier architecture allowing multiple levels of masters, it can
 be easily adapted to a wide range of virtual clusters distributed
 over geographically separated domains and heterogeneous machines
 (GRID's).

 The \Cmp{PROOF} technology has also proven to be very efficient in
 exploiting all the CPU's provided by many-core processors. A
 dedicated version of \Cmp{PROOF}, \Cmp{PROOF-Lite}, provides an
 out-of-the-box solution to take full advantage of the additional
 cores available in today desktops or laptops.

 Apart from the pure interactive mode, \Cmp{PROOF} has also an
 interactive-batch mode. With interactive-batch the user can start
 very long running queries, disconnect the client and at any time, any
 location and from any computer reconnect to the query to monitor its
 progress or retrieve the possibly intermediate results.
 This feature gives it a distinct
 advantage over purely batch based solutions, that only provide an
 answer once all sub-jobs have been finished and merged.


\subsubsection{PROOF Architecture}

\begin{figure}[t!]
  \centering
  \includegraphics[scale=0.4]{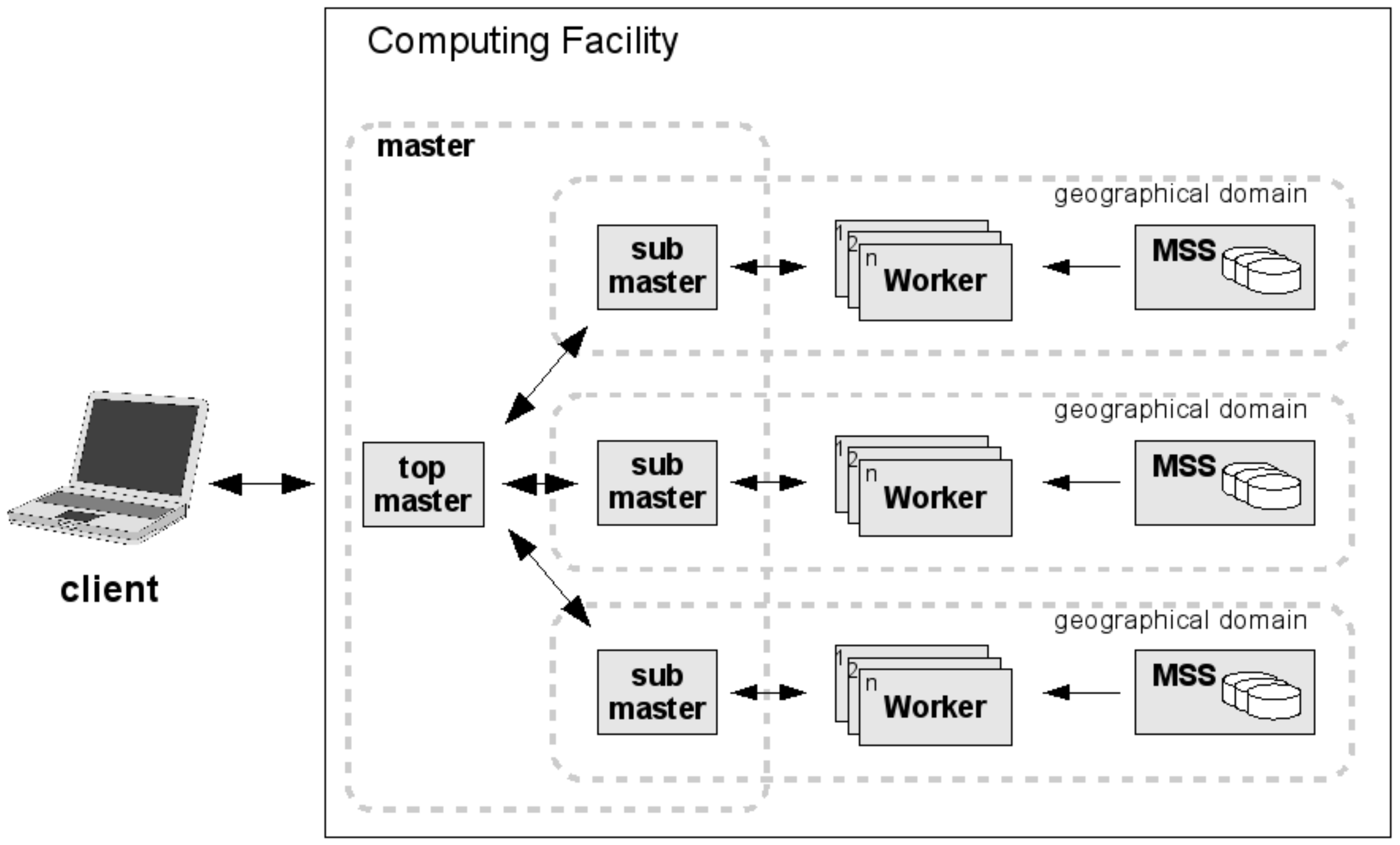}
  \caption{PROOF multi-tier master-worker architecture.}\label{fig-proof-arch}
\end{figure}

 The \Cmp{PROOF} system is implemented using a multi-tier architecture
 as shown in figure~\ref{fig-proof-arch}.

 The {\em client} is the user that wants to use the resources to
 perform a task. The {\em master} is the entry point to the computing
 facility: it parses the client requests, it distributes the work to
 the {\em workers}, it collects and merges the results. The master
 tier can be multi-layered. This allows, for example, to federate
 geographically separated clusters by optimizing the access to
 auxiliary resources, like mass storage systems (MSS). It also allows to
 distribute the distribution and merging work, which could otherwise become the
 bottle-neck in the case of many workers.

 \Cmp{PROOF-Lite}, the version of \Cmp{PROOF} dedicated to multicore
 desktops, implements a two-tier architecture where the master is
 merged into the client, the latter being in direct control of the
 workers.


\subsubsection{Event Level Parallelism}

 One of the ideas behind \Cmp{PROOF} is to minimize the execution
 time by having all contributing workers terminating their assigned
 tasks at the same time. This is achieved by using fine-grained work
 distribution, where the amount of work assigned to a worker,
 is adapted dynamically following the real-time
 performance of each worker. In principle, the packet can be as small
 as the basic unit, the event.

\begin{figure}[t!]
  \centering
  \includegraphics[scale=0.4]{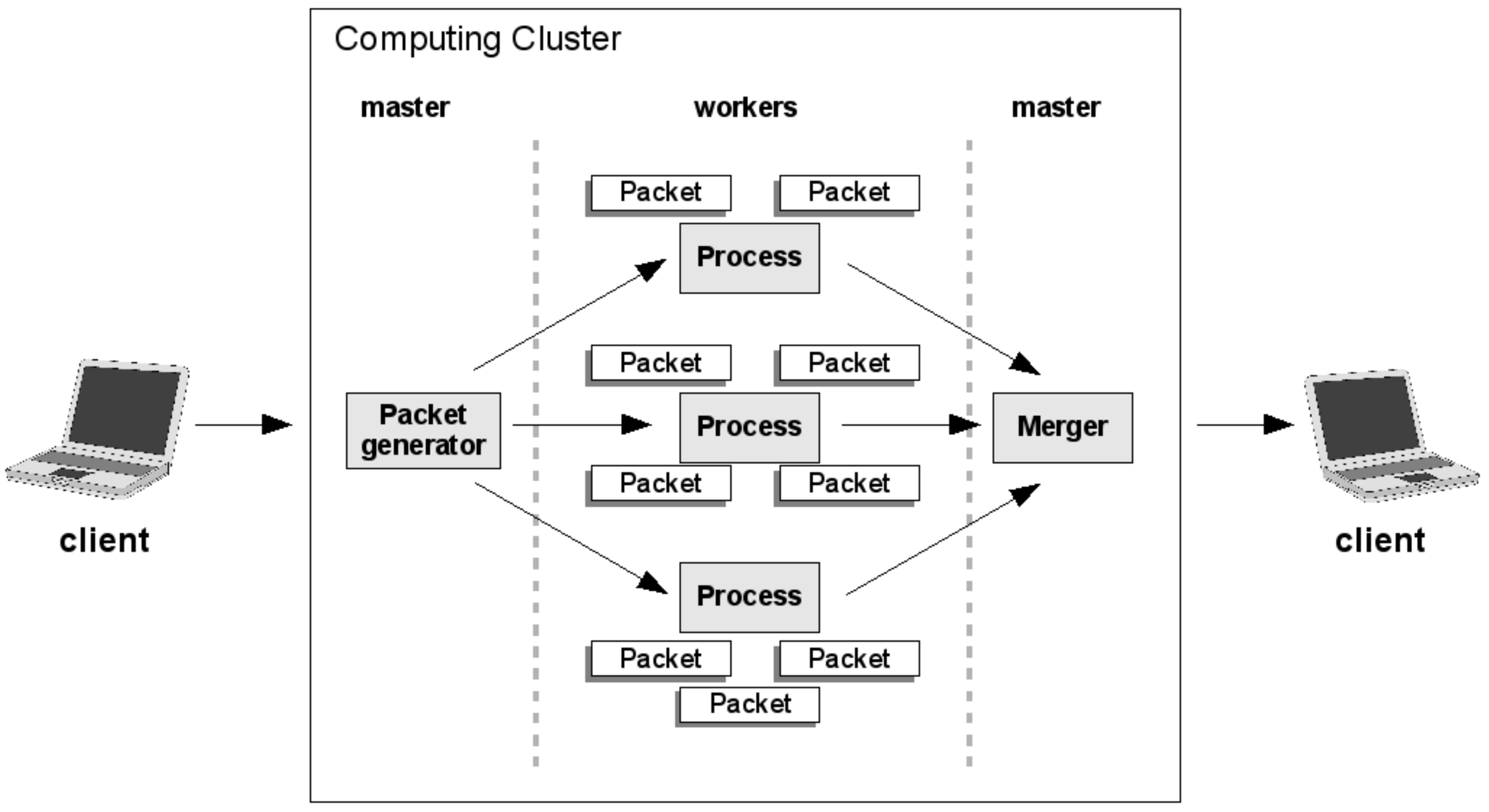}
  \caption{Schematic view of the PROOF workflow.}\label{fig-event-level-par}
\end{figure}

 A schematic view of the execution flow is given in
 figure~\ref{fig-event-level-par}.

\subsubsection{The Packetizer}

 The {\em packetizer} is responsible for load balancing a job between
 the workers assigned to it. It decides where each piece of work -
 called {\em a packet} - should be processed. An instance of the
 packetizer is created on the master node. In case of a multi-master
 configuration, there is one packetizer created for each of the
 sub-masters.

\begin{figure}[t!]
  \centering
  \includegraphics[scale=0.4]{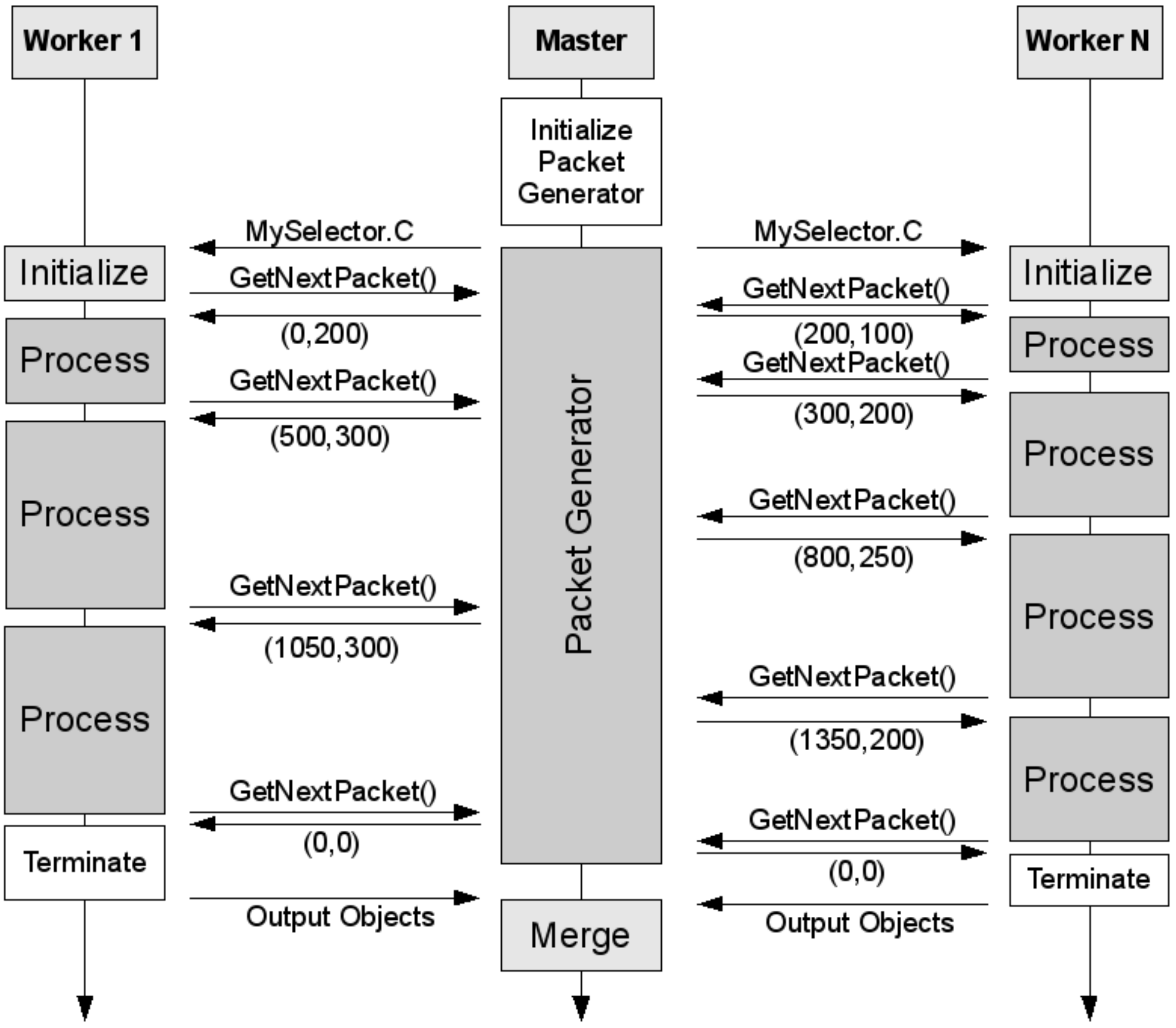}
  \caption{The PROOF {\em packetizer} distributes the work.}\label{fig-packetizer}
\end{figure}

 The performance of the workers can vary significantly as well as the
 file data transfer rates (local or remote files). In order to dynamically
 balance the work distribution, PROOF uses a ``pull
 architecture": when workers are ready for further processing they ask
 the packetizer for a next packet, see figure~\ref{fig-packetizer}.
 The packetizer uses a worker's processing rate to
 determine the size of the next packet for that worker. The packetizer tries to size
 all packets such that all workers will end at about the same time. At the
 beginning of a query the packets will be small, to quickly get an idea
 of the performance of the workers. Then the packet size will be increased
 to allow optimal disk access patterns (avoiding small reads) and to best
 suite the workers CPU performance.

\subsubsection{The Selector Framework}\label{sec-selector}

 To be able to perform event-level parallelism, \Cmp{PROOF} needs to
 be in charge of the event-loop, i.e.\ the execution flow steering the
 job. This requires that the code to be executed must have a
 predefined, though flexible structure. In {\ROOT} this is provided by
 the Selector framework, defined by the abstract class
 \Cls{TSelector}, which defines three logical steps:
\begin{enumerate}
\item \Mtd{Begin}, where the job definition (parameters, input data,
      outputs) is given; executed on the client and the workers;
\item \Mtd{Process}, where the actual job is done; called for each
      event, on the workers;
\item \Mtd{Terminate}, where the results are finally manipulated
      (fitted, visualized, etc.); called on the client and the
      workers.
\end{enumerate}

 \Mtd{Process} is the part that can be parallelized for the class of
 problems addressed by \Cmp{PROOF}.


\subsubsection{Aggregation of Results}

 \Cmp{PROOF} has a powerful feature that complements the use of the  \Cls{TSelector} framework.
After each worker has executed the  \Mtd{Terminate} method described above,
it sends the set of named results back to its master. The master collects these intermediate results
and aggregates them depending on their type. For several common types, like for example histograms,
there is a natural way to combine these results. The histogram obtained by adding all intermediate histograms
together is identical to the one that would have resulted from a single worker processing all events.
Similarly, event lists can be aggregated etc. \Cmp{PROOF} uses a well defined API for this process
allowing user defined classes to make use of this feature.
Intermediate results that cannot be combined are returned "as is" in a single collection for each resulting object.


\subsubsection{Real Time Monitoring and Feedback}

The user can monitor the progress of a \Cmp{PROOF} query or job in a number of different ways. 
A widget shows the number of events and files processed, the \% completed and the estimated time to completion.
This feedback is useful to get a high level idea of the behavior and performance of the \Cmp{PROOF} system and
its underlying components.

If the user registered histograms in the \Mtd{Begin} method of the \Cls{TSelector} class, \Cmp{PROOF} can show
these histograms, updating dynamically, during the running of the query. This feature allows the progress of the query to be monitored
in detail, especially if a very large data-set is being processed. The dynamically updating display is also very effective
in educational and demonstration settings.


\section{Installation Instructions}

 {\ROOT} can be build from source on all supported platforms using the
 well known Open Source tools like Subversion, \texttt{configure} and
 \texttt{make}.


\subsection{Getting the Source}

 The {\ROOT} source tar-ball can be obtained via ftp:

\begin{quote}
\begin{verbatim}
$ ftp root.cern.ch
User: anonymous
Password: <your-email-address>
ftp> cd root
ftp> bin
ftp> get root_<version>.source.tar.gz
ftp> bye
gzip -dc root_<version>.source.tar.gz | tar -xf -
\end{verbatim}
\end{quote}

 Alternatively the source can be obtained directly from the public
 Subversion repository:

\begin{quote}
\begin{verbatim}
svn co http://root.cern.ch/svn/root/trunk root
\end{verbatim}
\end{quote}

 A specific tag can be obtained using:

\begin{quote}
\begin{verbatim}
svn co http://root.cern.ch/svn/root/tags/v5-24-00 root-52400
\end{verbatim}
\end{quote}


\subsection{Compiling}

 Compiling {\ROOT} is just a matter of:

\begin{quote}
\begin{verbatim}
$ ./configure
$ make
\end{verbatim}
\end{quote}

 The \texttt{./configure} script will discover the platform and check for
 the existence of third party libraries needed for a number of
 optional plugins. To see all available options do:

\begin{quote}
\begin{verbatim}
$ ./configure --help
\end{verbatim}
\end{quote}

 For a complete description of the build procedure see the ROOT web site.


\section{Test Run Description}

 After installing {\ROOT} one can find a large set of test programs in
 the \texttt{tutorials} and \texttt{test} directories. The test
 programs in the \texttt{tutorials} directory are all in the form of
 macro's that can be either run via the \Cmp{CINT} interpreter or
 compiled via {\ACLiC}. A standard test macro is
 \texttt{benchmarks.C} that can be run via:

\begin{quote}
\begin{verbatim}
$ cd tutorials
$ root
root [0] .x benchmarks.C
root [1] .q
\end{verbatim}
\end{quote}

 If {\ROOT} is properly installed this macro should finish without
 errors and report a ROOTMARKS number:

\begin{quote}
\begin{verbatim}
****************************************************
* Your machine is estimated at 1120.18 ROOTMARKS   *
****************************************************
\end{verbatim}
\end{quote}

 The programs in the \texttt{test} directory are all stand-alone programs
 that are build by running \texttt{make}, like:

\begin{quote}
\begin{verbatim}
$ cd test
$ make
\end{verbatim}
\end{quote}

 This will compile a number of ``demo'' programs like, \texttt{guitest},
 \texttt{threads}, etc.\ and ``stress'' programs, like \texttt{stress}, {\tt
 stressGeometry}, \texttt{stressGraphics}, etc. All ``stress'' programs
 will also return a performance ROOTMARKS number, like:

\begin{quote}
\begin{verbatim}
$ ./stress -b 30
...
...
******************************************************************
*  ROOTMARKS = 859.2   *  Root5.23/03   20090226/1824
******************************************************************
\end{verbatim}
\end{quote}


\section{Acknowledgements}

 The success of an open source project can be measured by the number
 of users, but even more by the number of users who have turned into
 developers participating in improving the system with new code and
 bug fixes. We would like to thank all these users-turned-developers
 who have helped making {\ROOT} the powerful tool it is today.



\begin{thebibliography}{99}

\bibitem{root96}
R. Brun and F. Rademakers,
``ROOT -- An Object Oriented Data Analysis Framework'',
Proceedings AIHENP'96 Workshop, Lausanne, Sep. 1996, 
Nuclear Instruments and Methods in Physics Research A 389 (1997) 81-86.
\url{http://root.cern.ch/}

\bibitem{usersguide}
The ROOT Team,
``ROOT User's Guide'',
\url{http://root.cern.ch/drupal/content/users-guide}.
part of the ROOT submission to the Computer Physics Communications Program Library.

\bibitem{geant4}
S. Agostinelli et al,
``Geant 4---A simulation toolkit'',
Nuclear Instruments and Methods in Physics Research A 506 (2003) 250;\\
J. Allison et al,
`` Geant4 developments and applications'',
IEEE Transactions on Nuclear Science 53 (2006) 270.

\bibitem{geant3}
``GEANT. Detector Description and Simulation Tool'',
CERN Program Library Long Writeup W5013, 1993.

\bibitem{fluka}
A. Fass\`o et al,
``FLUKA: a multi-particle transport code'',
CERN-2005-10 (2005), INFN/TC\_05/11, SLAC-R-773;\\
G. Battistoni et al,
``The FLUKA code: Description and benchmarking'',
Proceedings of the Hadronic Shower Simulation Workshop 2006,
Fermilab 6--8 September 2006, M. Albrow, R. Raja eds.,
AIP Conference Proceeding 896 (2007) 31-49.

\bibitem{xrootd}
The Scalla/xrootd Team,
\emph{The Scalla Software Suite: xrootd/cmsd},
\url{http://xrootd.slac.stanford.edu/}

\bibitem{gccxml}
B.~King et al.,
\emph{GCC-XML, the XML Output Extension to GCC},
\url{http://www.gccxml.org}

\bibitem{dme}
L.~Janyst and R.~Brun and Ph.~Canal,
``ROOT Data Model Evolution'',
\url{http://root.cern.ch/root/SchemaEvolution.pdf}

\bibitem{ref}
The ROOT Development Team,
\emph{ROOT Reference Guide},
\url{http://root.cern.ch/root/html/}

\bibitem{zip}
D.A. Huffman,
``A Method for the Construction of Minimum-Redundancy Codes'',
Proceedings of the I.R.E., September 1952, pp 1098-1102.

\bibitem{MersenneTwister}
M. Matsumoto and T. Nishimura, 
``Mersenne twister: A 623-dimensionally equidistributed uniform 
pseudorandom number generato'',
 ACM Trans. on Modeling and Computer Simulations, 8, 1, (1998) 3-20.

\bibitem{gsl}
M. Galassi et al,
\emph{GNU Scientific Library Reference Manual}, third edition (January 2009),
\url{http://www.gnu.org/software/gsl/}

\bibitem{minuit}
F. James,
\emph{MINUIT.  Function Minimization and Error Analysis.  Reference Manual}
CERN Program Library Long Writeup D506,
\url{http://wwwasdoc.web.cern.ch/wwwasdoc/minuit/minmain.html}

\bibitem{minuit2}
M. Hatlo et al., 
``Developments of Mathematical Software Libraries for the LHC experiments'',
IEEE Transactions on Nuclear Science 52-6 (2005) 2818.

\bibitem{fumili}
S. Yashchenko,
``New method for minimizing regular functions with constraints on
parameter region'', 
Proceedings of CHEP'97 (1997)

\bibitem{lma}
K. Levenberg,
``A Method for the Solution of Certain Non-Linear Problems in Least Squares'',
The Quarterly of Applied Mathematics 2 (1944) 164--168.\\
D. Marquardt,
``An Algorithm for Least-Squares Estimation of Nonlinear Parameters'',
SIAM Journal on Applied Mathematics 11 (1963) 431--441.

\bibitem{smatrix} T. Glebe, ``SMatrix - A high performance library for Vector/Matrix calculation and Vertexing'', 
HERA-B Software Note 01-134, December 2, 2003.

\bibitem{mathChep07}
L. Moneta et al., ``Recent Developments of the ROOT Mathematical and Statistical Software'', Journal of Physics: 
Conference Series 119 (2008) 042023. 

\bibitem{foam}
S. Jadach, Computer Physics Communications 152 (2003) 55.

\bibitem{tmva}
A. H\"ocker et al., 
``TMVA - Toolkit for Multivariate Data Analysis'', CERN-OPEN-2007-007 
(2007), arXiv:physics/0703039v4
\url{http://tmva.sourceforge.net/}. 

\bibitem{roofit}W. Verkerke and D. Kirkby, 
``The RooFit Toolkit for data modeling'', proceedings to PHYSTAT05,
\url{http://roofit.sourceforge.net}.

\bibitem{rpp08}
C.~Amsler et al., 
``The Review of Particle Physics'',
Physics Letters B667 (2008) 1.

\bibitem{pyroot}
J.~Generowicz et al.,
``Reflection-Based Python--C++ Bindings",
LBNL Papers,
\url{http://repositories.cdlib.org/lbnl/LBNL-56538/}.

\bibitem{llvm}
Ch.~Lattner and V.~Adve,
``LLVM: A Compilation Framework for Lifelong Program Analysis \&
Transformation'',
Proceedings of the 2004 International Symposium on Code Generation and
Optimization (CGO04),
\url{http://www.llvm.org}

\bibitem{interp}
A.~Naumann and Ph.~Canal,
``The Role of Interpreters in High Performance Computing'',
Proceedings of ACAT 2008, PoS(ACAT08)065,
\url{http://pos.sissa.it/archive/conferences/070/065/ACAT08\_065.pdf}

\bibitem{proof}
M.~Ballintijn, M.~Biskup, R.~Brun, P.~Canal, D.~Feichtinger, G.~Ganis,
G.~Kickinger, A.~Peters, A.~Rademakers, 
``Parallel Interactive Data Analysis With PROOF'',
Nuclear Instruments and Methods in Physics Research
A 559 (2006) 13-16.

\bibitem{monarc}
M.~Aderholz et al.,
``Models of Networked Analysis at Regional Centres for LHC Experiments (MONARC)'',
CERN/LCB 2000-001,
\url{http://monarc.web.cern.ch/MONARC/}

\bibitem{aliroot}
Alice Collaboration,
``Technical Design Report of the Computing''
CERN-LHCC-2005-018, ALICE TDR 012
\url{http://aliceinfo.cern.ch/Offline/}

\bibitem{fairroot}
M. Al-Turany, F. Uhlig,
``FairRoot Framework'',
Proceedings of ACAT 2008, PoS(ACAT08)048,
\url{http://fairroot.gsi.de}

\end{thebibliography}
\end{document}